\documentclass[aps,prd,preprint,eqsecnum,tightenlines,nofootinbib]{revtex4}

\usepackage[dvips]{graphicx}
\usepackage{subfigure}
\usepackage{amsfonts}
\usepackage{amsmath, amsthm, amssymb}
\newif\ifdraft
\drafttrue
\newif\ifpreprint
\preprinttrue

\def\fig#1{Fig.~{\ref{#1}}}

\def\eqn#1{Eq.~({\ref{#1}})}
 
\def\sect#1{Section~{\ref{#1}}}

\def\app#1{Appendix~{\ref{#1}}}

\def\NeqFour{\mathcal{N}=4}

\def\I{{\cal I}}

\def\P{{\rm P}}
\def\NP{{\rm NP}}

\def\mud{\lambda}

\def\pol{\varepsilon}

\def\Tr{{\rm Tr}}

\def\spa#1.#2{\left\langle#1\,#2\right\rangle}
\def\spb#1.#2{\left[#1\,#2\right]}
\def\EE{{\cal E}}
\def\KE{{\cal K}}
\def\PE{{\cal P}}

\newbox\charbox
\newbox\slabox
\def\s#1{{      
        \setbox\charbox=\hbox{$#1$}
        \setbox\slabox=\hbox{$/$}
        \dimen\charbox=\ht\slabox
        \advance\dimen\charbox by -\dp\slabox
        \advance\dimen\charbox by -\ht\charbox
        \advance\dimen\charbox by \dp\charbox
        \divide\dimen\charbox by 2
        \raise-\dimen\charbox\hbox to \wd\charbox{\hss/\hss}
        \llap{$#1$}
}}

\newskip\humongous \humongous=0pt plus 1000pt minus 100pt

\newif\ifdtup

\newcounter{eqnumber}[section]

\begin{document}

\title{
\ifpreprint
 \hbox{\normalsize \rm MADCAP-13-06  \hskip 10.2 cm   UCLA/13/TEP/102} 
\hbox{$\null$}
\fi
\Large Color-Kinematics Duality for Pure Yang-Mills and Gravity at One
and Two Loops
}
 
\author{Zvi~Bern$^a$, Scott~Davies$^a$, Tristan~Dennen$^{b}$,
 Yu-tin~Huang$^{c}$, and Josh~Nohle$^a$}

\affiliation{
$a$ Department of Physics and Astronomy, University of California 
at Los Angeles\\ 
 Los Angeles, CA 90095-1547, USA \\ 
$\null$ \\
$b$ Niels Bohr International Academy and Discovery Center\\
The Niels Bohr Institute\\
Blegdamsvej 17, DK-2100 Copenhagen, Denmark\\
$\null$ \\
$c$ Michigan Center for Theoretical Physics \\
Randall Laboratory of Physics\\
University of Michigan, Ann Arbor, MI 48109, USA
$\null$ \\
}

\vskip .5 cm
\begin{abstract}
We provide evidence in favor of the conjectured duality
between color and kinematics for the case of nonsupersymmetric pure
Yang-Mills amplitudes by constructing a form of the one-loop
four-point amplitude of this theory that makes the duality manifest.
Our construction is valid in any dimension.  We also describe a
duality-satisfying representation for the two-loop four-point
amplitude with identical four-dimensional external helicities.  We use
these results to obtain corresponding gravity integrands
for a theory containing a graviton, dilaton, and antisymmetric tensor,
simply by replacing color factors with specified diagram numerators.
Using this, we give explicit forms of ultraviolet divergences at one
loop in four, six, and eight dimensions, and at two loops in four
dimensions.
\end{abstract}


\maketitle

\section{Introduction}

Recent years have seen remarkable progress in computing and
understanding scattering processes in gauge and gravity theories, both
for phenomenological and theoretical applications. (For various
reviews see Refs.~\cite{ReviewArticles,JJHenrikReview}.)  In
particular, various new structures have been uncovered in the
amplitudes of these theories (see, for example,
Ref.~\cite{VariousStructures}).  One such structure is the duality
between color and kinematics found by Carrasco, Johansson, and one of
the authors~\cite{BCJ, BCJLoop}.  This Bern-Carrasco-Johansson (BCJ) duality is conjectured to
hold at all loop orders in Yang-Mills theory and its supersymmetric
counterparts.  Besides imposing strong constraints on gauge-theory
amplitudes, whenever a form of a gauge-theory loop integrand is
obtained where the duality is manifest, we obtain corresponding
gravity integrands simply by replacing color factors by specified
gauge-theory kinematic numerator factors.

The duality between color and kinematics has been confirmed in
numerous tree-level
studies~\cite{Tye,StringBCJ,YMSquared,TreeBCJConf,OConnell,JJ2},
including the construction of explicit representations for an
arbitrary number of external legs~\cite{TreeAllN}.  At loop level, the
duality remains a conjecture, but there is already significant
nontrivial evidence in its favor for supersymmetric
theories~\cite{BCJLoop,ck4l,OneTwoLoopN4,SchnitzerBCJ,
  White,OneLoopN1Susy} and for special helicity configurations in
nonsupersymmetric pure Yang-Mills
theory~\cite{BCJLoop,OConnellRational}.  Here we provide further
evidence in favor of the duality at loop level, explicitly showing
that it holds for pure Yang-Mills one-loop four-point amplitudes for
all polarization states in $D$ dimensions.  We also present a duality-satisfying
representation of the two-loop four-point identical-helicity amplitude of pure
Yang-Mills.  This amplitude in a non-duality-satisfying representation was first
given in Ref.~\cite{Millenium}, while Ref.~\cite{BCJLoop} noted the existence
of a duality-satisfying form.  Here we explicitly give the full 
duality-satisfying form, including contributions from diagrams absent from
Ref.~\cite{Millenium} that vanish under integration but are necessary to
make the duality manifest.

In order to construct the one-loop four-point pure Yang-Mills
amplitude, we use a $D$-dimensional variant~\cite{DDimUnitarity} of
the unitarity method~\cite{UnitarityMethod}.  Our construction begins
by finding an ansatz for the amplitude constrained to satisfy the
duality.  Since the amplitude is fully determined from its
$D$-dimensional unitarity cuts, we obtain a form of the amplitude with
the duality manifest by enforcing that the ansatz has the correct
unitarity cuts.  The existence of such a form where both the duality
and the cuts are simultaneously satisfied is rather nontrivial.  We do
not use helicity states tied to specific dimensions but instead use
formal polarization vectors because we wish to have an expression for
the amplitude valid in any dimension and for all states.  The price for
this generality is that the expressions are lengthier.  Since the
constructed integrand has manifest BCJ duality, the double-copy
construction immediately gives the corresponding gravity amplitude in
a theory with a graviton, dilaton, and antisymmetric tensor.

We use these results to study the ultraviolet divergences of the
corresponding gravity amplitudes.  Recent years have seen a
renaissance in the study of ultraviolet divergences in gravity
theories, in a large measure due to the greatly improved ability to
carry out explicit multiloop computations in gravity
theories~\cite{LowerLoopSupergravity,BCJLoop, OneTwoLoopN4,
  SchnitzerBCJ, ck4l,ThreeloopHalfMax, TwoloopHalfMax}.  The unitarity
method also has revealed hints that multiloop supergravity theories
may be better behaved in the ultraviolet than power-counting arguments
based on standard symmetries suggest~\cite{Finite}.  Even pure
Einstein gravity at one loop exhibits surprising cancellations as the
number of external legs increases~\cite{UnexpectedOneLoop}.  The
question of whether it is possible to construct a finite supergravity
is still an open one, though there has been enormous progress on this
question in recent years, including new computations and a much better
understanding of the consequences of supersymmetry and duality
symmetry (see e.g. Refs.~\cite{SevenLoopE7,VanishingVolume}).  In
half-maximal supergravity~\cite{N4Sugra}, two- and three-loop examples
are known where the divergence vanishes, yet the understanding of the
possible symmetry behind this vanishing is
incomplete~\cite{VanishingVolume, ThreeloopHalfMax, VanhoveN4,
  TwoloopHalfMax}.  The duality between color and kinematics and its
associated double-copy formula offer a new angle on the ultraviolet
divergences in supergravity theories~\cite{BCJLoop, ck4l,
  ThreeloopHalfMax, TwoloopHalfMax,BoelsUV}.  Here we explore the
ultraviolet properties of nonsupersymmetric gravity from the
double-copy perspective.

We use the gravity integrands constructed via the double-copy property
to determine the exact form of the ultraviolet divergences.
  We do so at one loop in dimensions $D=4,6,8$.  The
ultraviolet properties of one-loop four-point gravity amplitudes have
already been studied in some detail over the years, including cases
with scalars or antisymmetric tensors coupling to
gravity~\cite{tHooftVeltman,OtherGravityMatter,antisymm,Deq6Div,
  DunbarNorridgeUV,UnexpectedOneLoop}, so no surprises should be
expected, at least at four points.  Nevertheless, it is useful to look
in some detail at the ultraviolet properties to understand
them from the double-copy perspective.  Here we
examine the four-point amplitudes in a theory of gravity coupled to a
dilaton and an antisymmetric tensor, corresponding to the double copy
of pure Yang-Mills theory.  While related calculations have been
carried out, we are unaware of any calculations of the ultraviolet
properties in the theory corresponding to the double-copy theory.

We find that in $D=4$, there are no one-loop divergences in amplitudes
involving external gravitons, though there are divergences in the
remaining amplitudes involving only external dilatons or antisymmetric
tensors, as expected from simple counterterm
arguments~\cite{tHooftVeltman}.  By two loops, even the four-graviton
amplitudes contain divergences, as we demonstrate by computing the
form and numerical coefficient of the divergence.  In the two-loop
case, the divergence is proportional to a unique $R^3$ operator which
gives a divergence in the identical-helicity four-point
amplitude. This means that the identical-helicity four-point amplitude
is sufficient for determining the coefficient of the $R^3$ divergence.
In $D=6$ and $D=8$, we find one-loop divergences in the
four-external-graviton amplitudes.  These results are not surprising
and are in line with the earlier studies.  Our conclusion is that, by
itself, the double-copy structure is insufficient to render a gravity
theory finite in $D=4$ and requires additional ultraviolet
cancellations, such as those from supersymmetry.

This paper is organized as follows.  In Section~\ref{sec:BCJReview},
we briefly review the duality between color and kinematics and the
double-copy construction of gravity.  In
Section~\ref{sec:Construction}, we present the construction of the
duality-satisfying pure Yang-Mills numerators at one and two loops.
Then in Section~\ref{sec:LoopUV}, we study the ultraviolet properties
of gravity coupled to a dilaton and an antisymmetric tensor at one
loop in four, six, and eight dimensions.  In the same section, we also
present the ultraviolet properties at two loops in four dimensions.
Finally, in Section~\ref{sec:Conclusion} we give our conclusions.
Appendices evaluating two-loop integrals needed in
Section~\ref{sec:twoLoopUV} are included.
Appendix~\ref{sec:DimRegAppendix} focuses on extracting the
divergences in dimensional regularization.  This procedure mixes
infrared and ultraviolet divergences; so, in
Appendix~\ref{sec:IRAppendix} we give the infrared divergences that
must be subtracted to obtain the ultraviolet ones.
Appendix~\ref{sec:UVIntegralsAppendix} evaluates the integrals
using an alternative method for obtaining the ultraviolet
divergences more directly, by introducing a mass to separate out the
infrared divergences from the ultraviolet ones.

\section{Review of BCJ Duality}
\label{sec:BCJReview}

An $L$-loop $m$-point gauge-theory amplitude in $D$ dimensions, with
all particles in the adjoint representation, may be written as
\begin{equation}
\mathcal{A}^{L\hbox{-}\mathrm{loop}}_{m}=i^{L} g^{m-2+2 L}\sum_{\mathcal{S}_{m}}\sum_{j}\int\prod_{l=1}^{L}\frac{d^{D}p_{l}}{(2\pi)^{D}}\frac{1}{S_{j}}\frac{c_{j} n_{j}}{\prod_{\alpha_{j}}p^{2}_{\alpha_{j}}}\,,
\label{CubicRepresentation}
\end{equation}
where $g$ is the gauge coupling constant. The first sum runs over the
$m!$ permutations of the external legs, denoted by $\mathcal{S}_{m}$.
The $S_{j}$ symmetry factor removes any overcounting from these
permutations and also from any internal automorphism symmetries of
graph $j$. The $j$-sum is over the set of distinct, nonisomorphic,
$m$-point $L$-loop graphs with only \emph{cubic} (i.e., trivalent)
vertices. These graphs are sufficient because any diagram with quartic
or higher vertices can be converted to a diagram with only cubic
vertices by multiplying and dividing by the appropriate propagators.
The propagators appearing in the graph are $1/\prod_{\alpha_{j}}
p^{2}_{\alpha_{j}}$.  The nontrivial kinematic information is
contained in the numerators $n_{j}$ and depends on momenta,
polarizations, and spinors.  In supersymmetric cases it will depend
also on Grassmann parameters, if a superspace form is used.  
The loop integral is over $L$ independent
$D$-dimensional loop momenta, $p_{l}$. Finally, $c_{j}$ denotes the
color factor, obtained by dressing every vertex in graph $j$ with the
group-theory structure constant,
$\tilde{f}^{abc}=i\sqrt{2}f^{abc}=\mathrm{Tr}([T^{a},T^{b}] T^{c})$,
where the hermitian generators of the gauge group are normalized via
$\mathrm{Tr}(T^{a}T^{b})=\delta^{a b}.$

\begin{figure}
\includegraphics[scale=.45]{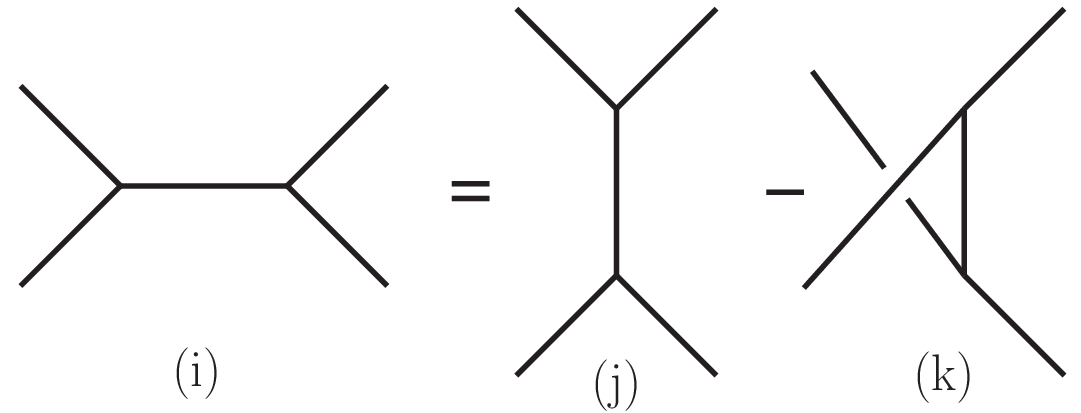}
\caption{The basic Jacobi relation for either color or numerator
  factors.  These three diagrams can be embedded in a larger diagram,
  including loops. }
\label{fig:BCJ}
\end{figure}

The numerators appearing in \eqn{CubicRepresentation} are by no means
unique because of freedom in moving terms between different diagrams.
Utilizing this freedom, the BCJ conjecture is that to all loop orders,
representations of the amplitude exist where kinematic numerators obey
the same algebraic relations that the color factors
obey~\cite{BCJ,BCJLoop}. In ordinary gauge theories, this 
is simply the Jacobi identity,
\begin{equation}
c_{i}=c_{j}-c_{k}\ \Rightarrow\ n_{i}=n_{j}-n_{k} \,,
\label{BCJDuality}
\end{equation}
where $i$, $j$, and $k$ label three diagrams whose color factors obey the
Jacobi identity.  The basic Jacobi identity is displayed
in \fig{fig:BCJ}.  The identity generalizes to any loop order with
any number of external legs by embedding it in larger diagrams, where
the other parts of the diagrams are identical for the three diagrams.
Furthermore, if the color factor of a diagram is antisymmetric under 
a swap of legs, we require that the numerator obey the same
antisymmetry,
\begin{equation}
c_{i} \rightarrow - c_{i}\ \Rightarrow\ n_{i} \rightarrow -n_{i} \,.
\label{BCJFlipSymmetry}
\end{equation}
The duality was noticed long ago for tree-level four-point Feynman
diagrams~\cite{Halzen}; beyond this, it is rather nontrivial 
and no longer holds for ordinary Feynman diagrams.
We note that the numerator relations are nontrivial functional
relations because they depend on momenta, polarizations, and spinors,
as discussed in some detail in Refs.~\cite{ck4l, JJHenrikReview,JJ2}.

While a complete understanding of the duality and its consequences is
still lacking, a variety of studies have elucidated it, especially at
tree level. In particular, this duality leads to nontrivial relations
between gauge-theory color-ordered partial tree
amplitudes~\cite{BCJ,BCJProofs}.  The duality (\ref{BCJFlipSymmetry})
has also been studied in string
theory~\cite{Bjerrum2,Tye,TreeBCJConf}.  In the self-dual case,
light-cone gauge Feynman rules have been shown to exhibit the
duality~\cite{OConnell}.  Explicit forms of $n$-point tree amplitudes
satisfying the duality have been found~\cite{TreeAllN}.  Although we
do not yet have a complete Lagrangian understanding, some progress in
this direction can be found in Refs.~\cite{YMSquared,OConnell}.  The
duality (\ref{BCJDuality}) does not need to be expressed in terms of
group structure constants but can alternatively be expressed in terms
of a trace-based representation~\cite{Trace}.  Progress has also been
made in understanding the underlying infinite-dimensional Lie
algebra~\cite{OConnell,OConnellAlgebras} responsible for the duality.
The duality between color and kinematics also appears to hold in
three-dimensional theories based on three algebras~\cite{BLGTheory},
as well as in some cases with higher-dimension
operators~\cite{BroedelDixon}.  Some initial studies of duality and its
implications for gravity in the high-energy limit have also been
carried out~\cite{HighEnergyLimit}.

At loop level, the duality remains a conjecture, but there is already
nontrivial evidence in its favor, especially for supersymmetric
theories.  At present, the list of loop-level cases where duality-satisfying forms of the amplitude are known to hold includes:
\begin{itemize}
	\item Up to four loops for four-point
         $\mathcal{N}=4$ super-Yang-Mills \cite{BCJLoop, ck4l} in a form valid 
         in $D$ dimensions;
	\item up to two loops for five external gluons in $\mathcal{N}=4$ 
         super-Yang-Mills theory~\cite{N4Five};
        \item up to seven points for one-loop amplitudes in $\mathcal{N}=4$ 
         super-Yang-Mills theory~\cite{Oxidation};
	\item up to two loops for four-point identical-helicity
         pure Yang-Mills amplitudes \cite{BCJLoop};
	\item through $n$ points for one-loop all-plus- or single-minus-helicity pure Yang-Mills amplitudes~\cite{OConnellRational};
        \item through four loops for a two-point (Sudakov) form factor in
          $\NeqFour$ super-Yang-Mills theory~\cite{BoelsFormFactor};
        \item one-loop four-point amplitudes in Yang-Mills theories with less
          than maximally supersymmetric amplitudes~\cite{OneLoopN1Susy}.
\end{itemize}
In this paper, we add the nonsupersymmetric pure Yang-Mills
one-loop four-point amplitude in $D$ dimensions to this list.  Besides direct
constructions, we note that the duality also appears to be consistent
with loop-level infrared properties of both gauge and gravity
theories~\cite{White}.

Another significant aspect of the duality is the ease with which
gravity loop integrands can be obtained from gauge-theory ones, once
the duality is made manifest~\cite{BCJ,BCJLoop}.  One simply replaces
the color factor with a kinematic numerator from a second gauge
theory,
\begin{equation}
c_{i}\ \rightarrow\ \tilde{n}_{i}\,.
\label{ColorSubstitution}
\end{equation}
This immediately gives the double-copy form of gravity amplitudes,
\begin{equation}
\mathcal{M}^{L\hbox{-}\mathrm{loop}}_{m} = i^{L+1} \left(\frac{\kappa}{2}\right)^{m-2+2 L}
\sum_{\mathcal{S}_{m}}\sum_{j}\int\prod_{l=1}^{L}\frac{d^{D}p_{l}}{(2\pi)^{D}}
\frac{1}{S_{j}}\frac{\tilde{n}_{j}n_{j}}{\prod_{\alpha_{j}}p^{2}_{\alpha_{j}}} \,,
\label{DoubleCopy}
\end{equation}
where $\tilde{n}_j$ and $n_j$ are gauge-theory numerator factors.  Only one of the
two sets of numerators needs to satisfy the duality
(\ref{BCJDuality})~\cite{BCJLoop,YMSquared} in order for the double-copy form (\ref{DoubleCopy}) to be valid.  The double-copy
formalism has been studied at loop level in some detail in a variety
of cases~\cite{BCJLoop,OneTwoLoopN4,N4Five,ck4l, ThreeloopHalfMax,
  TwoloopHalfMax, White, Oxidation}.

\section{Construction of Duality-Satisfying Integrands}
\label{sec:Construction}

We now describe the construction of a duality-satisfying
representation of the one-loop four-point amplitude in pure
Yang-Mills.  Since we want the form to be valid in all dimensions and
for all $D-2$ gluon states, we use formal polarizations instead of
helicity states.  This complicates the expression for the amplitude,
but has the advantage that it allows us to straightforwardly study the
amplitude and its gravity double copy in various dimensions.  In this
section, we also present a form of the two-loop pure Yang-Mills
identical-helicity amplitude given in Ref.~\cite{Millenium} that satisfies
BCJ duality after some rearrangement and addition of diagrams that
integrate to zero.  In
\sect{sec:twoLoopUV}, we use this amplitude to show that although
four-graviton amplitudes are ultraviolet finite in $D=4$ at one loop, they
diverge at two loops, in accordance with expectations.

\subsection{One Loop}

\begin{figure}
\includegraphics[scale=.75]{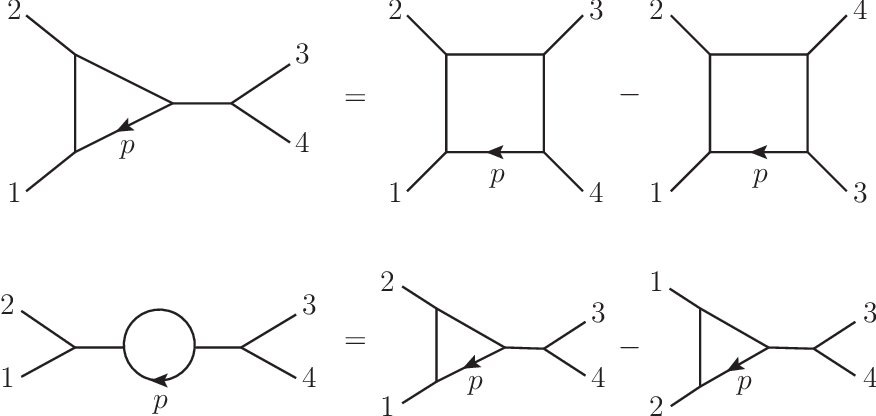}
\caption{The Jacobi relations determining either color or kinematic 
numerators of the four-point diagrams containing either a triangle  
or internal bubble. }
\label{fig:BCJOneLoop}
\end{figure}

\begin{figure}
\includegraphics[scale=.745]{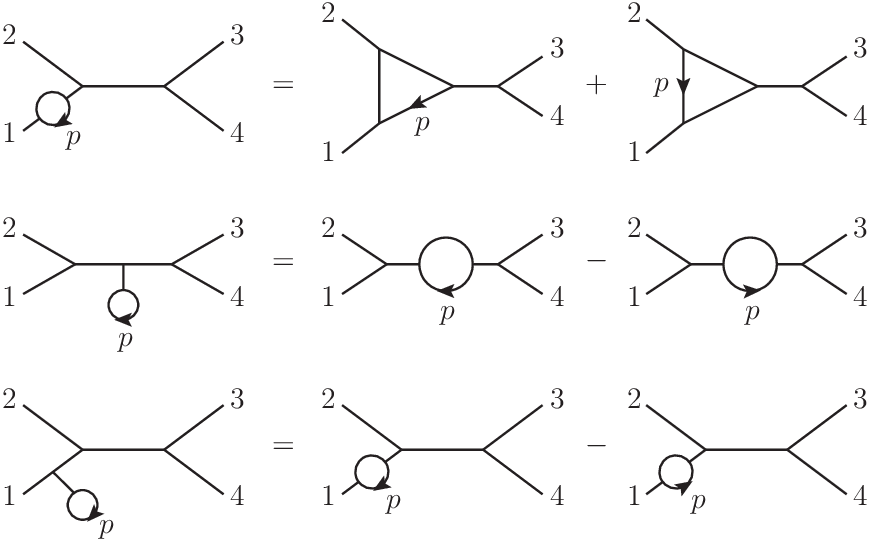}
\caption{The color or kinematic Jacobi relations involving a bubble
  on an external leg or a tadpole. These diagrams have vanishing
  contribution to the integrated amplitude.}
\label{fig:SnailTadpole}
\end{figure}

For a one-loop $n$-point amplitude, the duality~\eqref{BCJDuality} can
be used to express kinematic numerators of any diagram directly in
terms of $n$-gon numerators.  In particular, for the four-point case
we have two basic relations determining triangle and bubble
contributions from box numerators as illustrated in \fig{fig:BCJOneLoop},
\begin{eqnarray}
n_{12(34);p} &=& n_{1234;p} - n_{1243;p}\,, \nonumber \\
n_{(12)(34);p} &=& n_{12(34);p} - n_{21(34);p} \,.
\label{eq:boxBCJ}
\end{eqnarray}
The labels $1,2,3,4$ refer to the momenta and states of each external
leg, while the label $p$ denotes the loop momentum of the leg
indicated in \fig{fig:BCJOneLoop}.  (The parentheses in the 
subscript of the numerators indicate which external legs are pinched 
off to form a tree attached to the loop.)  Note that in
the figure the momentum of each internal leg of each diagram is the same
as in the other two diagrams \emph{except} for the single internal leg 
that differs between the diagrams.  In general, the bubble and triangle
contributions are nonvanishing; indeed, this explicitly holds for the
BCJ representation of the one-loop four-point amplitude of pure
Yang-Mills theory that we construct. 

Besides the diagrams in \fig{fig:BCJOneLoop}, there are diagrams with
a bubble on an external leg and diagrams with a tadpole, as shown in
\fig{fig:SnailTadpole}.  The duality also determines the numerators of
these diagrams via
\begin{eqnarray}
n_{1(234);p} &=& n_{12(34);p} + n_{1(43)2;p} \,, \nonumber \\
n_{(1234);p} &=& n_{(12)(34);p} - n_{(12)(34);-p} \,, \nonumber \\
n_{(\hat 1234);p} &=& n_{1(234);p} - n_{1(234);-p} \,,
\label{eq:snailBCJ}
\end{eqnarray}
corresponding respectively to the three relations in
\fig{fig:SnailTadpole}.  (On the final line in \eqn{eq:snailBCJ}, the
hat marks leg 1 as the location where the tadpole is attached.)  We
use these equations to impose the auxiliary constraint that the
tadpole numerators determined by BCJ duality vanish identically and
that all terms in the bubble-on-external-leg diagrams integrate to
zero as they do for Feynman diagrams.  Thus, these diagrams are not
necessary for determining the integrated amplitudes (though in $D=4$
the bubble-on-external-leg diagrams do affect the Yang-Mills
ultraviolet divergence).

Once we impose the BCJ conditions, the amplitude is entirely specified
by the box numerators.  Our task is then to find an expression for the
box numerators such that we obtain the correct amplitude.  It is
useful to impose a few auxiliary constraints to help simplify the
one-loop construction:
\begin{enumerate}
\item The box diagrams should have no more than four powers of loop
  momenta in the pure Yang-Mills case, matching the usual power count
  of Feynman-gauge Feynman diagrams.

\item Each numerator written in terms of formal polarization vectors
  respects the symmetries of the diagrams.  In
  particular, this condition implies that once a box diagram with one
  ordering of external legs is specified, the other orderings are obtained
  simply by relabeling.

\item The numerators of tadpole diagrams vanish prior to integration.

\item All terms in the bubble-on-external-leg diagrams integrate to
  zero, as they do for Feynman diagrams.  
\end{enumerate}
While it is not necessary to impose these conditions, they greatly
simplify the construction.  They ensure that the type of terms that
appear in the ansatz are similar to those of ordinary Feynman-gauge
Feynman diagrams, avoiding unnecessarily complicated terms.  (Using
generalized gauge invariance, one can always introduce arbitrarily
complicated terms into amplitudes, which cancel at the end.)

The first three conditions simplify the construction by restricting
the number of terms that appear.  The purpose of the fourth auxiliary
constraint is a bit more subtle.  While bubble-on-external-leg Feynman
diagrams are well defined in the on-shell limit, the freedom to
reassign terms used in the construction of BCJ numerators can
introduce ill-defined terms into such diagrams.  As a simple example,
consider the effect of the term $(k_1 + k_2)^2 \pol_1 \cdot k_2 \pol_2
\cdot k_1 \pol_3 \cdot \pol_4$ when added to the numerator of the
first diagram of \fig{fig:SnailTadpole} (with $k_i$ and $\pol_i$
external momenta and polarizations). Even after integration, this
contribution to the diagram is ill-defined because of the on-shell
intermediate propagator.  Such singular contributions would need to be
regularized by an appropriate off-shell continuation to ensure that
the introduced singularities cancel properly against singularities of
other diagrams.  While in principle we can introduce such a regulator,
it is best to avoid this complication altogether.  The fourth
condition ensures that we can treat the bubble-on-external-leg
contributions in the same way as for Feynman diagrams.  In particular,
with the constraint imposed, the bubble-on-external-leg contributions
match the Feynman-diagram property that they are proportional to
$(k_i^2)^{(D-4)/2}$, after accounting for the intermediate on-shell
propagator, and hence vanish in $D>4$, for $k_i$ on shell.  We note
that even with the fourth constraint, near $D=4$ we encounter the same
subtlety encountered with Feynman diagrams: Although
bubble-on-external-leg contributions are set to zero in dimensional
regularization, they can carry ultraviolet divergences.  Such
ultraviolet divergences cancel against infrared ones leaving a
vanishing result for on-shell bubble-on-external-leg diagrams.  The
net effect is that in gauge theory, we need to account for such
contributions to obtain the correct ultraviolet divergences.  In
contrast, in gravity even near $D=4$ there are neither infrared nor
ultraviolet divergences hiding in the bubble-on-external-leg
contributions because an extra two powers of numerator momenta give
rise to an additional vanishing.

We start the construction with an ansatz containing all possible
products of $\pol_i \cdot \pol_j$, $p\cdot\pol_i$, $k_i\cdot\pol_j$,
 $p\cdot k_i$, $p\cdot p$, $s$, and $t$, where the $k_i$ are three
independent external momenta, $p$ is the loop momentum, $\pol_i$ are
external polarization vectors, and
\begin{equation}
s=(k_1+k_2)^2\,, \hspace{1.5cm} t=(k_2+k_3)^2\,,
\end{equation} 
are the usual Mandelstam invariants.
By dimensional analysis, each
numerator term must contain four momenta in addition to being linear
in all four $\pol_i$'s. We also set $k_{i}\cdot\varepsilon_{i}=0$ and
impose momentum conservation with $k_{4}=-k_{1}-k_{2}-k_{3}$ and
$k_{1}\cdot\varepsilon_{4}=-k_{2}\cdot\varepsilon_{4}-k_{3}\cdot\varepsilon_{4}$.
This yields 468 terms, each with a coefficient to be determined. 

Our first constraint on the coefficients comes from demanding that the
box numerator obey the rotation and reflection symmetries of the box
diagram. This leaves us with 81 free
coefficients.  An ansatz for the full amplitude is then obtained by
using the duality relations~\eqref{eq:boxBCJ},~\eqref{eq:snailBCJ} to
determine numerators for all other diagrams.

\renewcommand{\thesubfigure}{(\arabic{subfigure})}

\subfigbottomskip = -0.2cm
\begin{figure}[ht]
\centering
\vskip .5 cm 
\subfigure{\includegraphics[scale=.7]{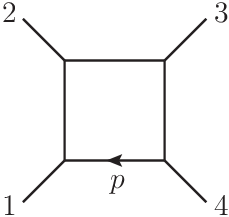}}
\hspace{1cm}
\subfigure{\includegraphics[scale=.7]{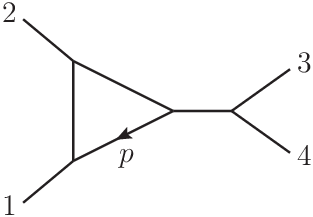}} 
\hspace{1cm}
\subfigure{\includegraphics[scale=.7]{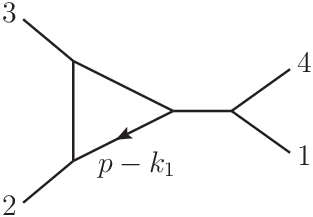}}\\
\vspace{.4cm}
\subfigure{\includegraphics[scale=.7]{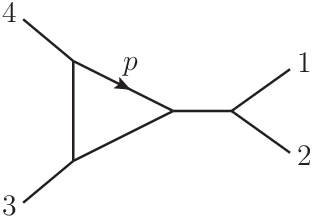}} 
\hspace{1.5cm} 
\subfigure{\includegraphics[scale=.7]{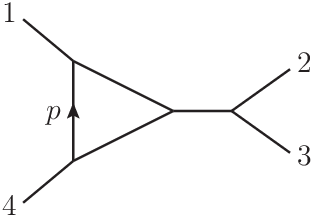}} \\ 
\vspace{.6cm} 
\subfigure{\includegraphics[scale=.7]{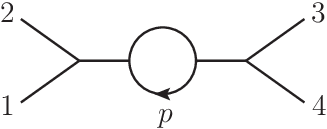}}
\hspace{1.5cm}
\subfigure{\includegraphics[scale=.7]{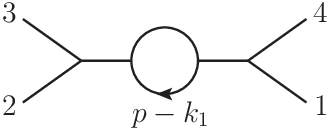}}
\vspace{.2cm}
\caption[a]{The seven diagrams for the color-ordered 
amplitude with ordering $(1,2,3,4)$.
}
\label{fig:ColorOrderDiags}
\end{figure}
\renewcommand{\thesubfigure}{(\alph{subfigure})}
\begin{figure}[ht]
\centering
\subfigure[]{\includegraphics[scale=.6]{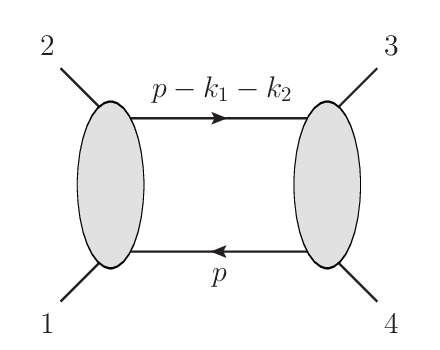}}
\hspace{1.5cm}
\subfigure[]{\includegraphics[scale=.6]{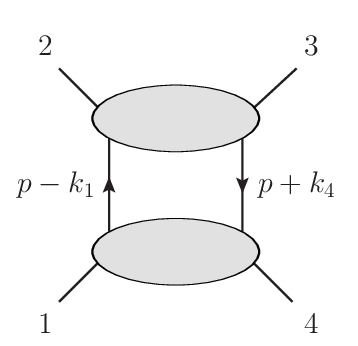}} 
\caption[a]{The (a) $s$-channel and (b) $t$-channel unitarity cuts
 used to determine the amplitude.  The exposed
intermediate legs are on shell.
 }
\label{fig:TwoParticleCut}
\end{figure} 
The next step is to determine coefficients in the ansatz by matching
to the unitarity cuts of the amplitude. It is convenient to use a
color-ordered form of the amplitude~\cite{BKColor} for this
matching. The seven diagrams contributing to the color-ordered
amplitude, that is the coefficient of the color trace
$N_c\Tr[T^{a_1}T^{a_2} T^{a_3} T^{a_4}]$, are shown in
\fig{fig:ColorOrderDiags}.  The other color-ordered amplitudes are
simple relabelings of this one.  For the one-loop four-point
amplitude, the $s$- and $t$-channel unitarity cuts---shown in
\fig{fig:TwoParticleCut}---are sufficient to determine this
color-ordered amplitude up to terms that integrate to zero.  One
straightforward means for determining the cuts is to construct the
amplitude in Feynman gauge and then take its unitarity cuts at the
integrand level prior to integration.  This automatically gives us an
expression for the cuts valid in $D$ dimensions without any spurious
denominators (such as light-cone denominators from physical state
projectors).  This matching procedure nontrivially rearranges the
amplitude so that BCJ duality is manifest.  After matching the cuts,
we also impose the fourth auxiliary condition to tame the
bubble-on-external-leg contributions.  Finally we impose that the
tadpole numerators vanish.  Including all the auxiliary constraints
with these conditions, we can solve for all but five free
coefficients.  Because the $s$- and $t$-channel unitarity cuts are
independent of these parameters, the integrated amplitude should not
depend on them.

Using the shorthand notation,
\begin{eqnarray}
&& p_1=p\,, \hspace{1.cm} p_2=p-k_1\,, \hspace{1.cm} p_3=p-k_1-k_2\,, 
\hspace{1.cm} p_4=p+k_4\,, \nonumber \\
&& \EE_{ij}=\varepsilon_i\cdot\varepsilon_j\,, \hspace{2cm} 
\PE_{ij}=p_i\cdot\varepsilon_j\,, \hspace{2cm} 
\KE_{ij} =k_i\cdot\varepsilon_j\,,
\end{eqnarray} 
and setting the free parameters to zero for simplicity, the box
numerator is 
\begin{align}
 \lefteqn{ \hskip -.5 cm 
n_{1234;p}\: =  -i \Bigl[
\tfrac{D_{s}-2}{8} \: \EE_{14} \: \EE_{23}   \: p_{1}^{2} \: p_{3}^{2}
+\tfrac{D_{s}-2}{24} \: \EE_{13} \: \EE_{24} \: p_{1}^{2} \: p_{3}^{2} 
-\tfrac{D_{s}-2}{24} \: \EE_{12} \: \EE_{34} \: p_{1}^{2} \: p_{3}^{2} 
-\tfrac{2}{3} \: \EE_{14} \: \EE_{23} \: p_{3}^{2} \: s}
\vspace{.2cm} \nonumber \\ 
& \null
-\tfrac{2}{3} \: \EE_{13} \: \EE_{24} \: p_{2}^{2} \: s
+\tfrac{2}{3} \: \EE_{12} \: \EE_{34} \: p_{2}^{2} \: s
+\tfrac{2}{3} \: \EE_{14} \: \EE_{23} \: p_{2}^{2} \: s
+\tfrac{1}{2} \: \EE_{14} \: \EE_{23} \: s^{2}
+ 2 \: \EE_{23} \: \KE_{24} \: \KE_{41} \: p_{3}^{2}
\vspace{.2cm} \nonumber \\
& \null
+\tfrac{D_{s}-74}{24} \: \EE_{13} \: \KE_{12} \: \KE_{34} \: p_{3}^{2}
+\tfrac{D_{s}-74}{24} \: \EE_{24} \: \KE_{23} \: \KE_{41} \: p_{3}^{2}
-\tfrac{D_{s}-26}{3} \: \EE_{12} \: \KE_{13} \: \KE_{34} \: p_{3}^{2}
\vspace{.2cm} \nonumber \\ 
& \null
-\tfrac{D_{s}-26}{6} \: \EE_{34} \: \KE_{41} \: \KE_{42} \: p_{3}^{2}
-\tfrac{D_{s}-26}{2} \: \EE_{12} \: \KE_{23} \: \KE_{34} \: p_{3}^{2}
-\tfrac{D_{s}-26}{2} \: \EE_{34} \: \KE_{12} \: \KE_{41} \: p_{3}^{2}
\vspace{.2cm} \nonumber \\ 
& \null
+\tfrac{D_{s}-26}{12} \: \EE_{34} \: \KE_{31} \: \KE_{42} \: p_{3}^{2}
+\tfrac{5(D_{s}-26)}{24} \: \EE_{24} \: \KE_{13} \: \KE_{41} \: p_{3}^{2}
-\tfrac{D_{s}-26}{8} \: \EE_{24} \: \KE_{13} \: \KE_{31} \: p_{3}^{2}
\vspace{.2cm} \nonumber \\ 
& \null
-\tfrac{11(D_{s}-26)}{24} \: \EE_{24} \: \KE_{23} \: \KE_{31} \: p_{3}^{2}
-\tfrac{D_{s}-26}{24} \: \EE_{34} \: \KE_{12} \: \KE_{31} \: p_{3}^{2}
+\tfrac{D_{s}-30}{2} \: \EE_{13} \: \KE_{12} \: \KE_{24} \: p_{3}^{2}
\vspace{.2cm} \nonumber \\ 
& \null
-\tfrac{D_{s}-14}{6} \: \EE_{13} \: \KE_{34} \: \KE_{42} \: p_{3}^{2}
+\tfrac{D_{s}-38}{6} \: \EE_{13} \: \KE_{24} \: \KE_{42} \: p_{3}^{2}
-\tfrac{5(D_{s}-26)}{24} \: \EE_{12} \: \KE_{13} \: \KE_{24} \: p_{3}^{2}
\vspace{.2cm} \nonumber \\ 
& \null
-\tfrac{11(D_{s}-26)}{24} \: \EE_{12} \: \KE_{23} \: \KE_{24} \: p_{3}^{2}
+\tfrac{13D_{s}-290}{24} \: \EE_{14} \: \KE_{23} \: \KE_{42} \: p_{3}^{2}
+(D_{s}\! -\!24) \: \EE_{14} \: \KE_{12} \: \KE_{23} \: p_{3}^{2}
\vspace{.2cm} \nonumber \\ 
& \null
+\tfrac{11(D_{s}-26)}{24} \: \EE_{14} \: \KE_{13} \: \KE_{42} \: p_{3}^{2}
+\tfrac{11(D_{s}-26)}{24} \: \EE_{14} \: \KE_{12} \: \KE_{13} \: p_{3}^{2}
-\tfrac{D_{s}-26}{12} \: \EE_{23} \: \KE_{24} \: \KE_{31} \: p_{3}^{2}
\vspace{.2cm} \nonumber \\ 
& \null
-\tfrac{D_{s}-26}{12}\: \EE_{23} \: \KE_{31} \: \KE_{34} \: p_{3}^{2}
-\tfrac{D_{s}-50}{12} \: \EE_{23} \: \KE_{34} \: \KE_{41} \: p_{3}^{2}
- 4 \: \EE_{14} \: \KE_{12} \: \KE_{23} \: s
- 2 \: \EE_{23} \: \KE_{24} \: \KE_{31} \: s
\vspace{.2cm} \nonumber \\ 
& \null
- 2 \: \EE_{23} \: \KE_{24} \: \KE_{41} \: s
- 2 \: \EE_{12} \: \KE_{23} \: \KE_{24} \: s
- 2 \: \EE_{14} \: \KE_{12} \: \KE_{13} \: s
- 2 \: \EE_{12} \: \KE_{23} \: \KE_{34} \: s
\vspace{.2cm} \nonumber \\ 
& \null
+\tfrac{7D_{s}-230}{12}\: \EE_{23} \: \KE_{31} \: \PE_{44} \: p_{3}^{2}
+\tfrac{7D_{s}-230}{24} \: \EE_{23} \: \KE_{34} \: \PE_{11} \: p_{3}^{2}
+\tfrac{7D_{s}-230}{24}\: \EE_{23} \: \KE_{41} \: \PE_{44} \: p_{3}^{2}
\vspace{.2cm} \nonumber \\ 
& \null
+\tfrac{7D_{s}-230}{24} \: \EE_{13} \: \KE_{34} \: \PE_{22} \: p_{3}^{2}
+\tfrac{7D_{s}-230}{24}\: \EE_{24} \: \KE_{41} \: \PE_{33} \: p_{3}^{2}
-\tfrac{7(D_{s}-26)}{24} \: \EE_{24} \: \KE_{13} \: \PE_{11} \: p_{3}^{2}
\vspace{.2cm} \nonumber \\ 
& \null
+\tfrac{7(D_{s}-26)}{24}\: \EE_{12} \: \KE_{13} \: \PE_{44} \: p_{3}^{2}
-\tfrac{7D_{s}-230}{24} \: \EE_{23} \: \KE_{24} \: \PE_{11} \: p_{3}^{2}
-\tfrac{7D_{s}-230}{24}\: \EE_{12} \: \KE_{24} \: \PE_{33} \: p_{3}^{2}
\vspace{.2cm} \nonumber \\ 
& \null
-\tfrac{7D_{s}-230}{24} \: \EE_{34} \: \KE_{42} \: \PE_{11} \: p_{3}^{2}
-\tfrac{11D_{s}-238}{24} \: \EE_{13} \: \KE_{12} \: \PE_{44} \: p_{3}^{2}
-\tfrac{11D_{s}-238}{24} \: \EE_{24} \: \KE_{23} \: \PE_{11} \: p_{3}^{2}
\vspace{.2cm} \nonumber \\ 
& \null
+ 2\: \EE_{12} \: \KE_{23} \: \PE_{44} \: p_{3}^{2}
+ 2 \: \EE_{34} \: \KE_{12} \: \PE_{11} \: p_{3}^{2}
-\tfrac{D_{s}-14}{6} \: \EE_{13} \: \KE_{42} \: \PE_{44} \: p_{3}^{2}
-\tfrac{3(D_{s}-26)}{8} \: \EE_{34} \: \KE_{31} \: \PE_{22} \: p_{3}^{2}
\vspace{.2cm} \nonumber \\ 
& \null
-\tfrac{3(D_{s}-26)}{8} \: \EE_{24} \: \KE_{31} \: \PE_{33} \: p_{3}^{2}
-\tfrac{2(D_{s}-29)}{3} \: \EE_{34} \: \KE_{41} \: \PE_{22} \: p_{3}^{2}
-\tfrac{2(D_{s}-29)}{3} \: \EE_{12} \: \KE_{34} \: \PE_{33} \: p_{3}^{2}
\vspace{.2cm} \nonumber \\ 
& \null
+\tfrac{13D_{s}-290}{24} \: \EE_{14} \: \KE_{42} \: \PE_{33} \: p_{3}^{2}
+\tfrac{13D_{s}-290}{24} \: \EE_{14} \: \KE_{12} \: \PE_{33} \: p_{3}^{2}
+\tfrac{13D_{s}-290}{24} \: \EE_{14} \: \KE_{23} \: \PE_{22} \: p_{3}^{2}
\vspace{.2cm} \nonumber \\ 
& \null
+\tfrac{2(D_{s}-29)}{3} \: \EE_{13} \: \KE_{24} \: \PE_{22} \: p_{3}^{2}
- 2 \: \EE_{14} \: \KE_{42} \: \PE_{33} \: s
- 2 \: \EE_{34} \: \KE_{41} \: \PE_{22} \: s
- 2 \: \EE_{14} \: \KE_{12} \: \PE_{33} \: s
\vspace{.2cm} \nonumber \\ 
& \null
- 2 \: \EE_{12} \: \KE_{24} \: \PE_{33} \: s
- 2 \: \EE_{12} \: \KE_{34} \: \PE_{33} \: s
- 2 \: \EE_{14} \: \KE_{23} \: \PE_{22} \: s
+ 2 \: \EE_{13} \: \KE_{34} \: \PE_{22} \: s
+ 2 \: \EE_{24} \: \KE_{41} \: \PE_{33} \: s
\vspace{.2cm} \nonumber \\ 
& \null
+ 2 \: \EE_{13} \: \KE_{24} \: \PE_{22} \: s
- (D_{s}\! -\!2)\: \EE_{23} \: \PE_{11} \: \PE_{44} \: p_{3}^{2}
- \tfrac{D_{s}-2}{6} \: \EE_{13} \: \PE_{22} \: \PE_{44} \: p_{3}^{2}
- \tfrac{D_{s}-2}{6} \: \EE_{24} \: \PE_{33} \: \PE_{11} \: p_{3}^{2}
\vspace{.2cm} \nonumber \\
& \null
+ \tfrac{D_{s}-2}{6} \: \EE_{12} \: \PE_{33} \: \PE_{44} \: p_{3}^{2}
+ \tfrac{D_{s}-2}{6} \: \EE_{34} \: \PE_{11} \: \PE_{22} \: p_{3}^{2}
- 4 \: \EE_{34} \: \PE_{11} \: \PE_{22} \: s
+ 2 \: \EE_{13} \: \PE_{22} \: \PE_{44} \: s
\vspace{.2cm} \nonumber \\ 
& \null
+ 2 \: \EE_{24} \: \PE_{33} \: \PE_{11} \: s
+ 4 \: \KE_{12} \: \KE_{13} \: \KE_{24} \: \KE_{31}
+ 4 \: \KE_{12} \: \KE_{23} \: \KE_{24} \: \KE_{31}
+ 2 \: \KE_{12} \: \KE_{13} \: \KE_{31} \: \KE_{34}
\vspace{.2cm} \nonumber \\ 
& \null
+ 4 \: \KE_{12} \: \KE_{23} \: \KE_{31} \: \KE_{34}
+  \: \KE_{13} \: \KE_{24} \: \KE_{31} \: \KE_{42}
+ 2 \: \KE_{12} \: \KE_{23} \: \KE_{34} \: \KE_{41}
- 4 \: \KE_{12} \: \KE_{24} \: \KE_{41} \: \PE_{33}
\vspace{.2cm} \nonumber \\ 
& \null
+ 4 \: \KE_{31} \: \KE_{34} \: \KE_{42} \: \PE_{33}
+ 4 \: \KE_{24} \: \KE_{41} \: \KE_{42} \: \PE_{33}
+ 4 \: \KE_{34} \: \KE_{41} \: \KE_{42} \: \PE_{33}
+ 4 \: \KE_{24} \: \KE_{31} \: \KE_{42} \: \PE_{33}
\vspace{.2cm} \nonumber \\
& \null
- 8 \: \KE_{34} \: \KE_{41} \: \PE_{22} \: \PE_{33} 
- 8 \: \KE_{24} \: \KE_{41} \: \PE_{22} \: \PE_{33}
+ 4 \: \KE_{24} \: \KE_{42} \: \PE_{11} \: \PE_{33}
\vspace{.2cm} \nonumber \\
& \null
+  (D_{s} \!-\! 2) \: \PE_{11} \: \PE_{22}\: \PE_{33}  \: \PE_{44}\Bigr]
+\mathrm{cyclic}\,,
\label{YMNum}
\end{align}
where $D_s$ is a state-counting parameter, so that $D_s-2$ is the
number of gluon states circulating in the loop.  The notation `$+$
cyclic' indicates that one should include the three additional cyclic
permutations of indices, giving a total of four permutations
$(1,2,3,4)$, $(2,3,4,1)$, $(3,4,1,2)$, $(4,1,2,3)$ of all variables
$\varepsilon_i,k_i,p_i,s,t$.  Plain-text, computer-readable versions
of the full expressions for the numerators, including also gluino- and
scalar-loop contributions, can be found online~\cite{Online}.  In
\eqn{YMNum}, we have written the expression for the box numerator in a
different form than that available online in order to exhibit the cyclic
symmetry.

We have explicitly checked that after reducing the pure Yang-Mills
amplitude to an integral basis,\footnote{We thank R. Roiban for
  cross-checking our computation.} the expression is free of arbitrary
parameters and in $D=4$ matches the known expression for the amplitude in
Ref.~\cite{BK}, after accounting for the fact that the expression in
that paper is renormalized.  The reduction for four-dimensional
external states was carried out by expanding the external
polarizations in terms of the external momenta plus a dual
vector~\cite{epMomBasis}.

As another simple cross check, we have extracted the ultraviolet
divergences in $D=6,8$ and compared them to the known forms.  In
$D=6,8$, with our fourth auxiliary constraint there are no ultraviolet
contributions from bubbles on external legs.  This allows us to
directly extract the ultraviolet divergences by introducing a mass
regulator and then expanding in small external momenta using the
methods of Ref.~\cite{smallMomenta}.  We find complete agreement with
both earlier evaluations in Ref.~\cite{venMetsaev}.
We have also compared this to an extraction of the ultraviolet
divergences directly using dimensional regularization without
introducing an additional mass regulator and again find agreement.

\subsection{Two Loops}
\label{sec:twoLoopBCJ}

We now turn to two loops. 
As we shall discuss in \sect{sec:LoopUV}, the four-graviton
amplitude in the double-copy theory is ultraviolet finite at
one loop. To test whether this continues at two loops, we need the
two-loop amplitude.  As it turns out, the identical-helicity amplitude
is sufficient for our purposes because the divergence comes from 
an $R^3$ operator whose coefficient is fixed by this amplitude. 
We therefore now turn to finding a
form of the two-loop identical-helicity amplitude where BCJ duality is
manifest.  It would be interesting to obtain a general two-loop
construction valid for all states in $D$ dimensions, but we do not do so here.

\begin{figure}[tb]
\def\scale{.52}
\centering
\subfigure[]{\includegraphics[scale=\scale]{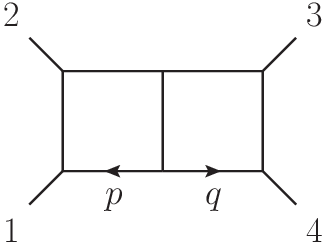}}
\hspace{1cm}
\subfigure[]{\includegraphics[scale=\scale]{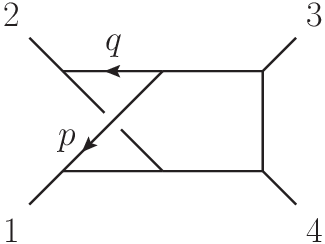}}
\hspace{1cm}
\subfigure[]{\includegraphics[scale=\scale]{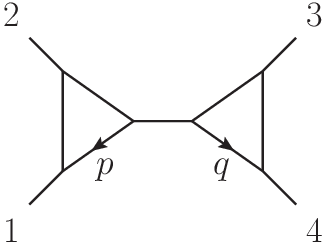}}
\hspace{1cm}
\subfigure[]{\includegraphics[scale=\scale]{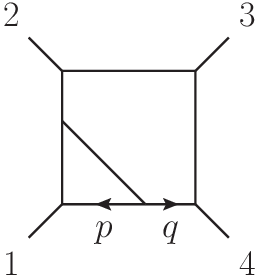}}
\\
\subfigure[]{\includegraphics[scale=\scale]{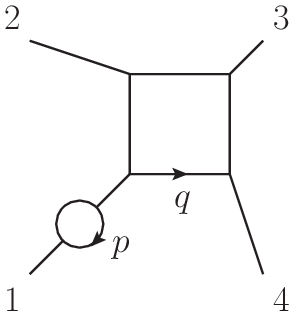}}
\hspace{1.2cm}
\subfigure[]{\includegraphics[scale=\scale]{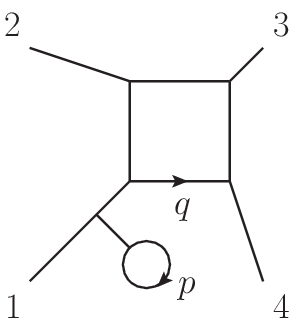}}
\hspace{1.2cm}
\subfigure[]{\includegraphics[scale=\scale]{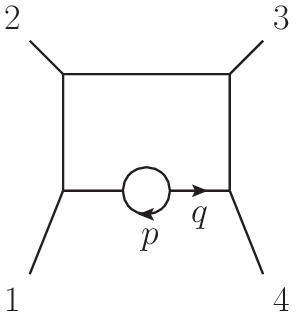}}
\hspace{1.2cm}
\subfigure[]{\includegraphics[scale=\scale]{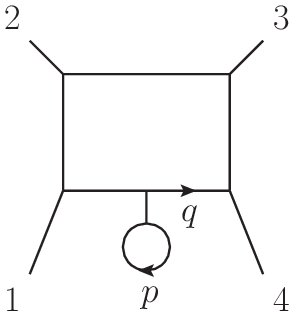}}
\\
\subfigure[]{\includegraphics[scale=\scale]{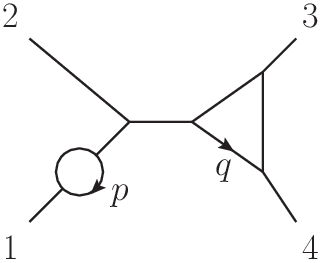}}
\hspace{1.2cm}
\subfigure[]{\includegraphics[scale=\scale]{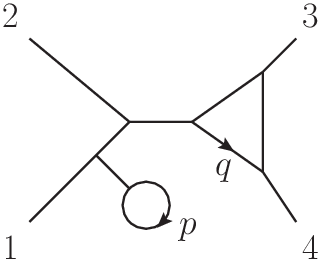}}
\hspace{1.2cm}
\subfigure[]{\includegraphics[scale=\scale]{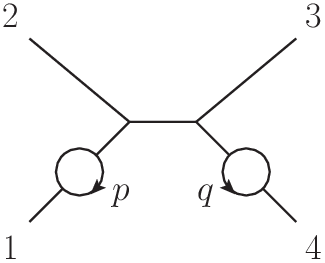}}
\caption[a]{The diagrams needed to describe an integrand for the
  identical helicity-amplitude where the duality between color and
  kinematics is manifest.  When integrated all diagrams, except the
  (a) planar double-box, (b) nonplanar double-box, and (c)
  double-triangle integrals, vanish.}
\label{Fig:PandNP}
\end{figure}

\begin{figure}[tb]
\centering
\hskip .01 cm 
\includegraphics[scale=.8]{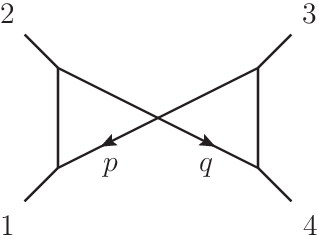}
\caption[a]{The bow-tie integral appearing
in the identical-helicity pure Yang-Mills amplitude. }
\label{Fig:bowtie}
\end{figure}

The identical-helicity pure Yang-Mills amplitude has previously been
constructed in Ref.~\cite{Millenium}.  There the amplitude is given in
the following representation:
\begin{align}
\mathcal{A}_4^{(2)}(1^+,2^+,3^+,4^+)=
g^6\,\frac{1}{4}\sum_{\mathcal{S}_4}\left[ 
C^{\mathrm{P}}_{1234}A^{\mathrm{P}'}_{1234}
+C^{\mathrm{NP}}_{12;34}A^{\mathrm{NP}}_{12;34}\right],
\label{eq:twoLoopYMOrig}
\end{align}
where the sum runs over all 24 permutations of the external legs. We will 
describe the all-plus-helicity case; the all-negative-helicity 
case follows from parity conjugation. The
prefactor of $1/4$ accounts for the overcount due to
symmetries of the diagrams.  $C^{\mathrm{P}}_{1234}$ and
$C^{\mathrm{NP}}_{12;34}$ are the color factors obtained from the
planar double-box and nonplanar double-box diagrams
shown in \fig{Fig:PandNP}(a) and (b), respectively, by dressing each
vertex with an $\tilde{f}^{abc}$ and summing over the contracted color
indices.  $A^{\mathrm{P}'}_{1234}$ and $A^{\mathrm{NP}}_{12;34}$ are then the associated partial amplitudes.
These partial amplitudes are~\cite{Millenium}
\begin{align}
A_{1234}^{\mathrm{P}'} & =
i\mathcal{T}\,
\biggl\{s\,\mathcal{I}_4^{\mathrm{P}}\bigl[(D_s-2)\left(\lambda_p^2\lambda_q^2+\lambda_p^2\lambda_{p+q}^2+ \lambda_q^2\lambda_{p+q}^2\right)+16\left((\lambda_p\cdot\lambda_q)^2-\lambda_p^2\lambda_q^2\right)\bigr]\!(s,t) \notag \\
&\hspace{2.2cm}+4(D_s-2)\mathcal{I}_4^{\mathrm{bow\hbox{-}tie}}\left[\left(\lambda_p^2+\lambda_q^2\right)\left(\lambda_p\cdot\lambda_q\right)\right]\!(s) \notag \\
&\hspace{2.2cm}+\frac{(D_s-2)^2}{s}\mathcal{I}_4^{\mathrm{bow\hbox{-}tie}}\left[\lambda_p^2\lambda_q^2((p+q)^2+s)\right]\!(s,t)\biggr\}\,, \notag \\
A_{12;34}^{\mathrm{NP}}& =
i\mathcal{T}\,
s\, \mathcal{I}_4^{\mathrm{NP}}\bigl[(D_s-2)\left(\lambda_p^2\lambda_q^2
  +\lambda_p^2\lambda_{p+q}^2 +
\lambda_q^2\lambda_{p+q}^2\right) 
+16\left((\lambda_p\cdot\lambda_q)^2-\lambda_p^2\lambda_q^2\right)\bigr](s,t)\,,
\label{eq:PNP}
\end{align}
where the permutation-invariant kinematic prefactor is given by
\begin{equation}
\mathcal{T}\equiv\frac{[1\,2][3\,4]}{\langle 1\,2\rangle\langle 3\,4\rangle}\,,
\end{equation}
where the angle and square brackets are standard spinor inner
products.  For the all negative-helicity case, the angle and
square products should be swapped.  The planar double-box
(Fig.~\ref{Fig:PandNP}(a)), nonplanar double-box
(Fig.~\ref{Fig:PandNP}(b)), and bow-tie integrals
(Fig.~\ref{Fig:bowtie}) are
\begin{align}
& \mathcal{I}_4^{\mathrm{P}}[\mathcal{P}(\lambda_i,p,q,k_i)](s,t) \nonumber \\
& \hskip 1 cm 
\equiv\int\frac{d^Dp}{(2\pi)^D}\frac{d^Dq}{(2\pi)^D}\frac{\mathcal{P}(\lambda_i,p,q,k_i)}{p^2q^2(p+q)^2(p-k_1)^2(p-k_1-k_2)^2(q-k_4)^2(q-k_3-k_4)^2}\,, \nonumber \\
&\mathcal{I}_4^{\mathrm{NP}}[\mathcal{P}(\lambda_i,p,q,k_i)](s,t) \nonumber \\
&\hskip 1 cm 
\equiv\int\frac{d^Dp}{(2\pi)^D}\frac{d^Dq}{(2\pi)^D}
\frac{\mathcal{P}(\lambda_i,p,q,k_i)}{p^2q^2(p+q)^2(p-k_1)^2(q-k_2)^2(p+q+k_3)^2(p+q+k_3+k_4)^2}\,, \nonumber \\
&\mathcal{I}_4^{\mathrm{bow\hbox{-}tie}}[\mathcal{P}(\lambda_i,p,q,k_i)](s,t) \nonumber \\
&\hskip 1 cm 
\equiv\int\frac{d^Dp}{(2\pi)^D}\frac{d^Dq}{(2\pi)^D}
\frac{\mathcal{P}(\lambda_i,p,q,k_i)}{p^2q^2(p-k_1)^2(p-k_1-k_2)^2 
(q-k_4)^2(q-k_3-k_4)^2}\,, 
\end{align}
where $\lambda_p$, $\lambda_q$, and $\lambda_{p+q}$ represent the 
$(-2\epsilon)$-dimensional components of loop momenta $p$, $q$, and $(p+q)$.

Ref.~\cite{BCJLoop} notes that
a representation where the numerators satisfy the BCJ duality can be
obtained directly from the representation of the amplitude given in
Ref.~\cite{Millenium}.  Here we describe this in more detail, including
additional diagrams that integrate to zero and are
undetectable in ordinary unitarity cuts, but are needed to make the
duality manifest.

\subfigbottomskip = -0.1cm
 \subfigcapskip = -.1cm

\subfigbottomskip = -0.1cm
 \subfigcapskip = -.1cm

\begin{figure}[tb]
\def\hs{\hspace{.6cm}}
\def\scale{.39}
\begin{center}
\subfigure[]{\includegraphics[scale=\scale]{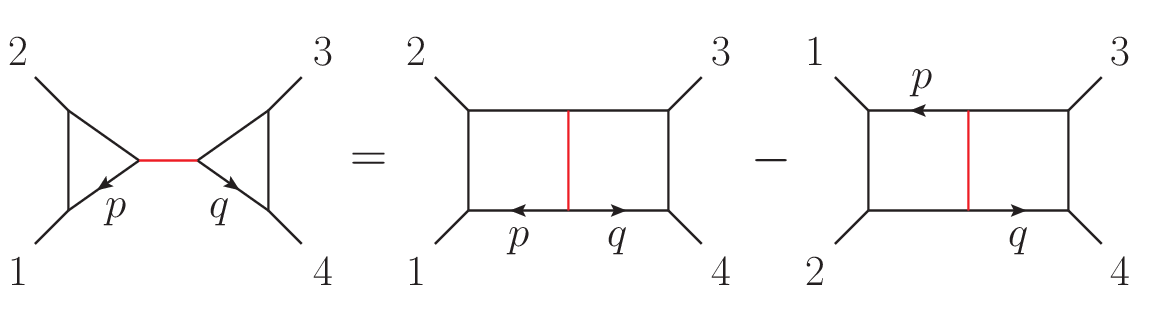}}
\hs
\subfigure[]{\includegraphics[scale=\scale]{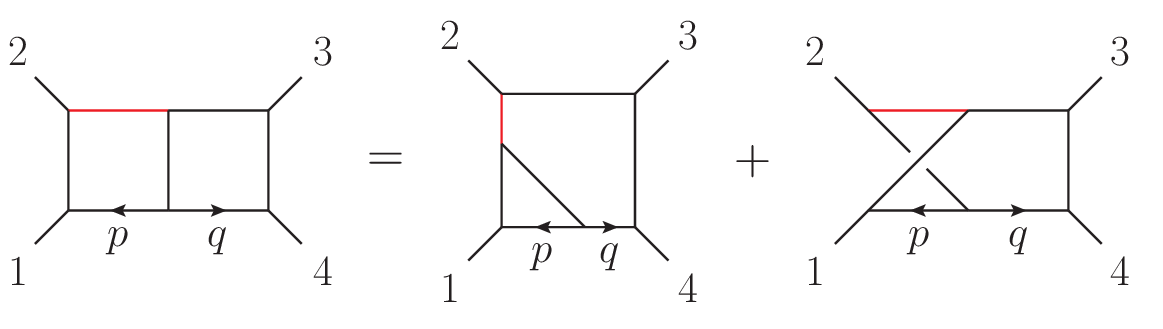}}
\\
\subfigure[]{\includegraphics[scale=\scale]{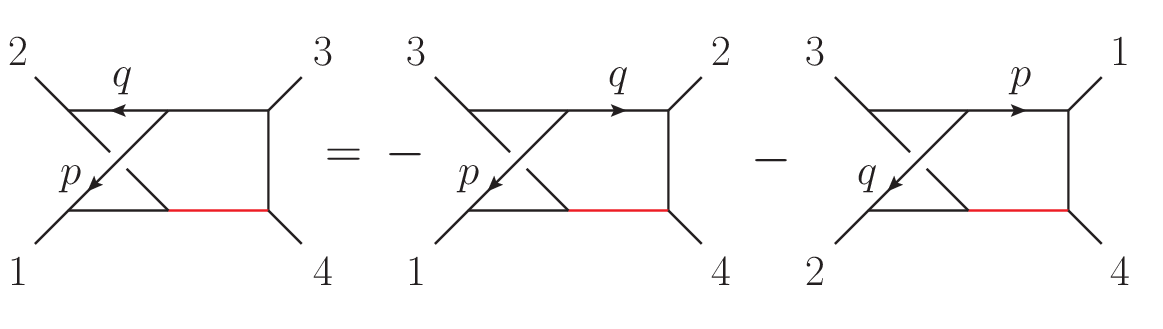}}
\hs
\subfigure[]{\includegraphics[scale=\scale]{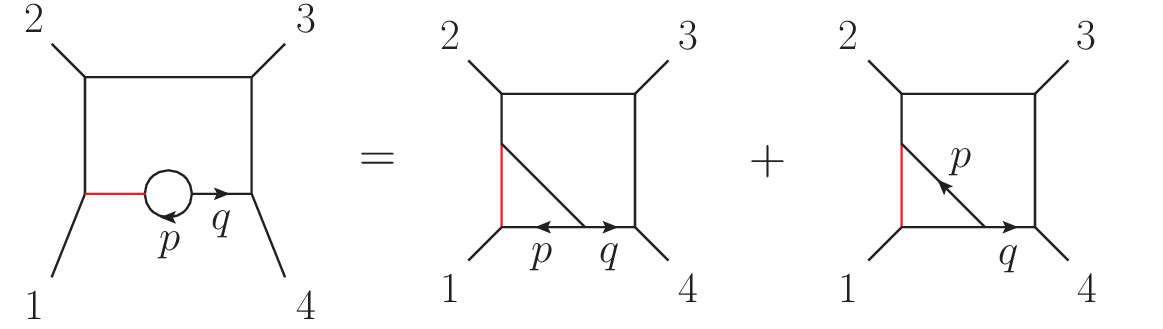}}
\\
\subfigure[]{\includegraphics[scale=\scale]{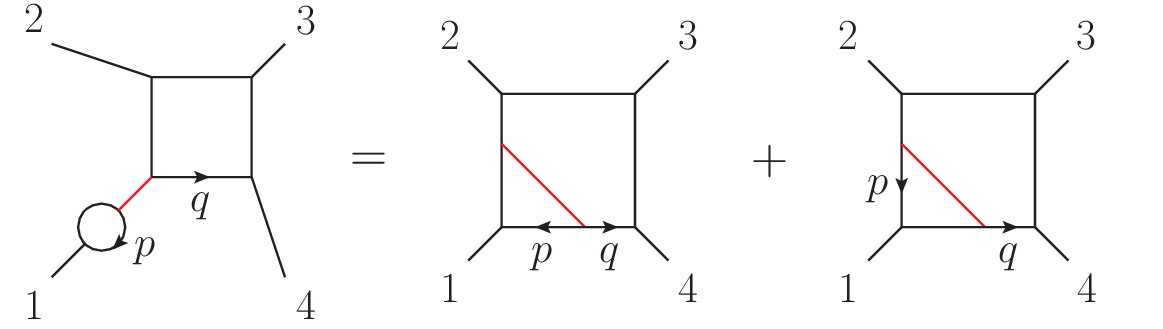}}
\hs
\subfigure[]{\includegraphics[scale=\scale]{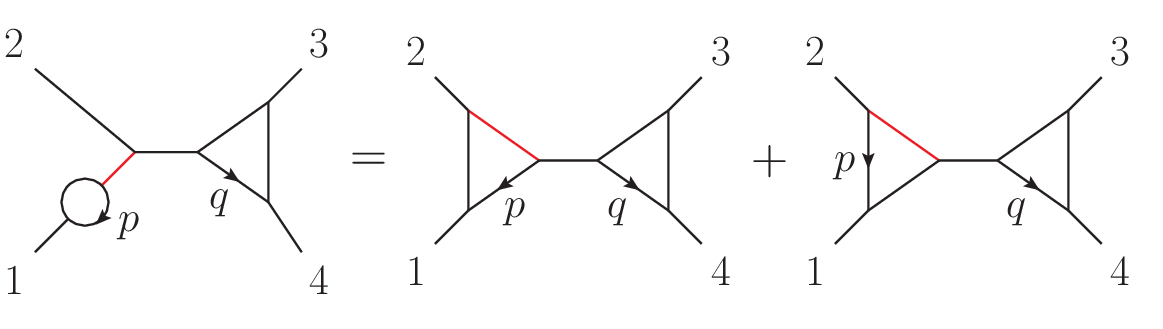}}
\\
\subfigure[]{\includegraphics[scale=\scale]{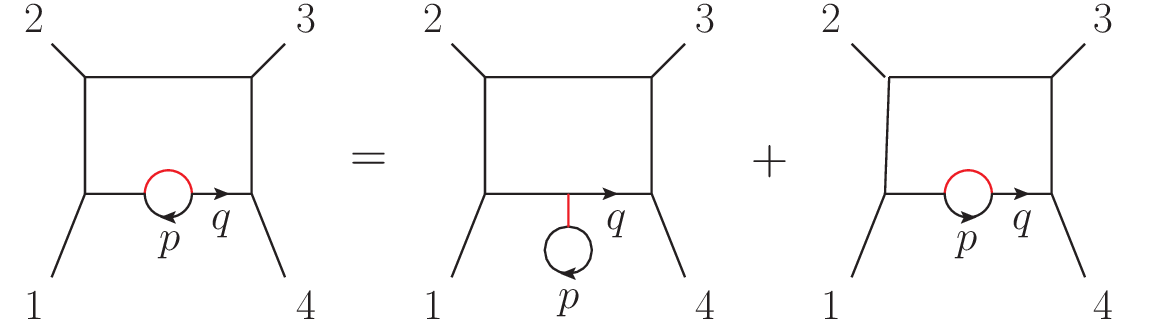}}
\hs
\subfigure[]{\includegraphics[scale=\scale]{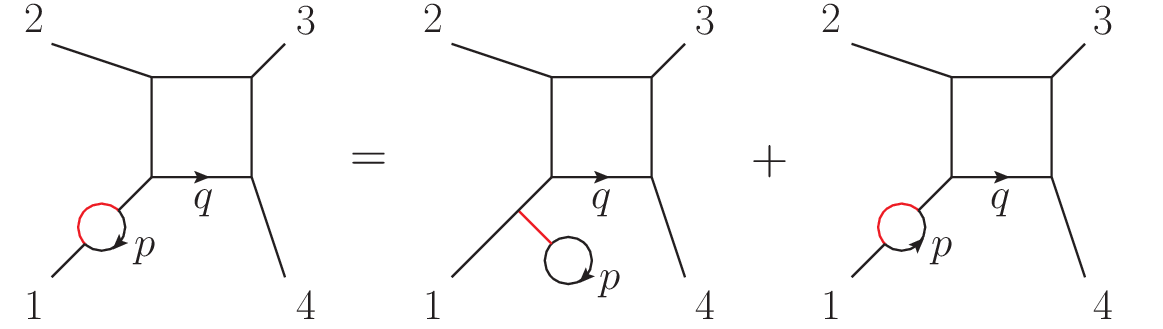}}
\\
\subfigure[]{\includegraphics[scale=\scale]{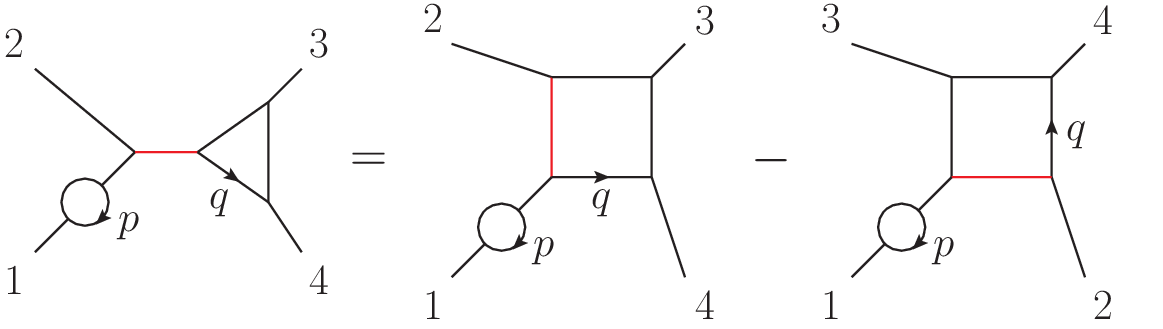}}
\hs
\subfigure[]{\includegraphics[scale=\scale]{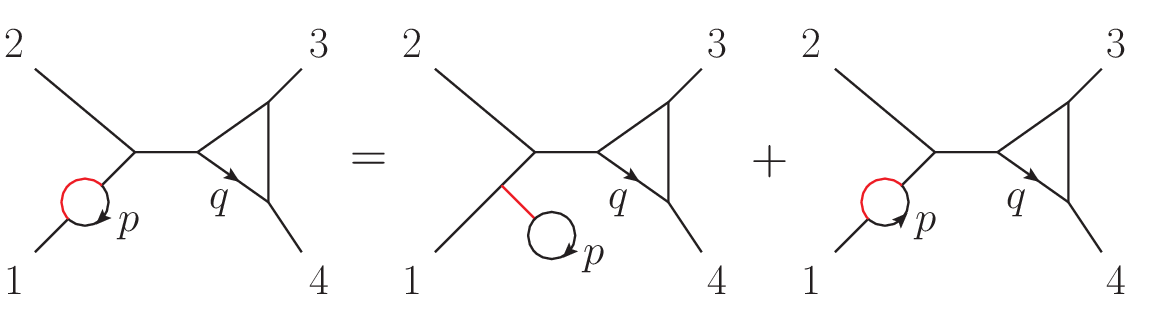}}
\\
\subfigure[]{\includegraphics[scale=\scale]{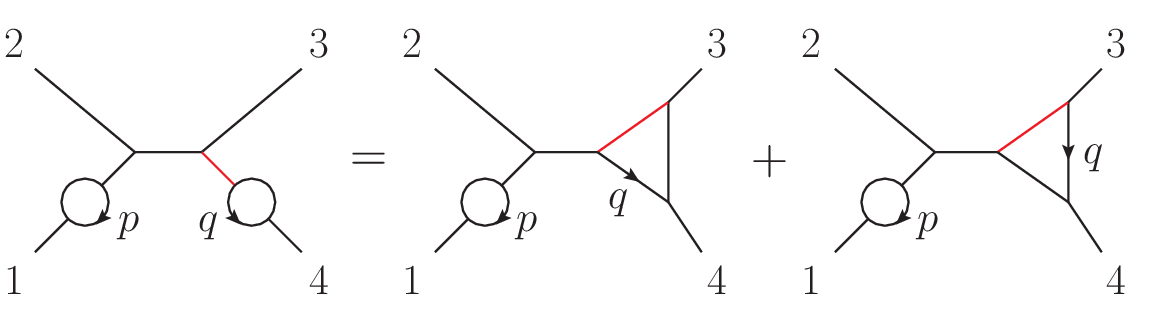}}
\hs
\subfigure[]{\includegraphics[scale=\scale]{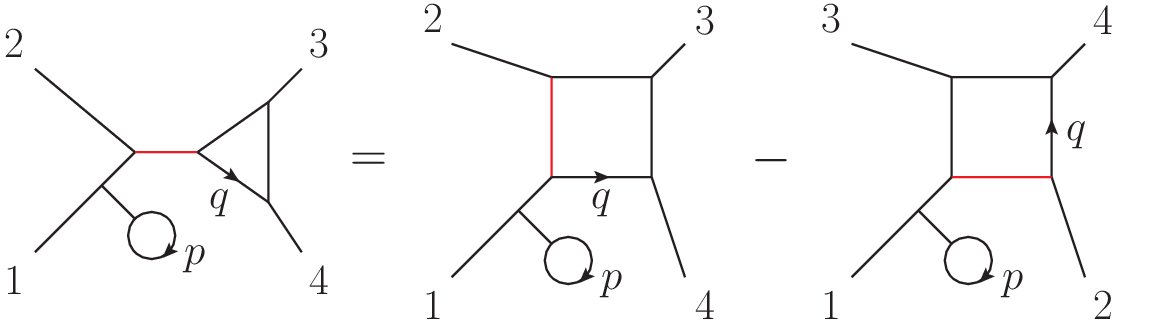}}
\\
\subfigure[]{\includegraphics[scale=\scale]{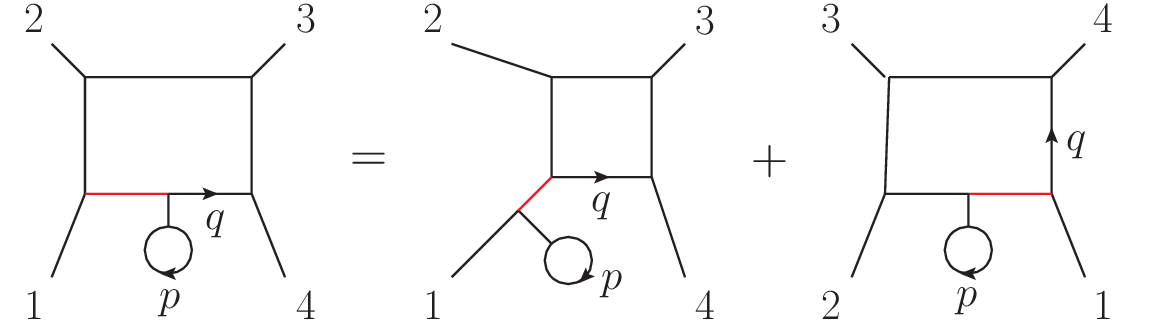}}
\caption{The nontrivial duality relations (a)-(m) satisfied by the numerators
  of the identical-helicity two-loop amplitude.  The shaded (red) leg
  marks the central leg of the applied Jacobi identity.}
\label{Fig:twoLoopBCJ}
\end{center}
\end{figure}

We begin with a rearranged form of the identical-helicity amplitude,
\begin{align}
\mathcal{A}_4^{(2)}(1^+,2^+,3^+,4^+)=
g^6\sum_{\mathcal{S}_4}\left[ 
  \frac{1}{4} C^{\mathrm{P}}_{1234}A^{\mathrm{P}}_{1234}
+ \frac{1}{4} C^{\mathrm{NP}}_{12;34}A^{\mathrm{NP}}_{12;34}
+ \frac{1}{8} C^{\mathrm{DT}}_{1234} A^{\mathrm{DT}}_{1234}\right].
\label{eq:twoLoopYM}
\end{align}
$C^{\mathrm{DT}}_{1234}$ is the color factor obtained from the stretched bow-tie or double-triangle diagram
in \fig{Fig:PandNP}(c).  $A^{\mathrm{NP}}_{12;34}$ is given in Eq.~\eqref{eq:PNP}, while $A^{\mathrm{P}}_{1234}$ and
$A^{\mathrm{DT}}_{1234}$ are
\begin{align}
A_{1234}^{\mathrm{P}} & =
i\mathcal{T}\,
\mathcal{I}_4^{\mathrm{P}}\biggl[
\frac{(D_s-2)^2}{2}(p+q)^2\lambda_p^2\lambda_q^2
+ 16s\left((\lambda_p\cdot\lambda_q)^2-\lambda_p^2\lambda_q^2\right) \nonumber \\
& \hskip 1 cm \null
+(D_s-2) \Bigl(s\left(\lambda_p^2\lambda_q^2+\lambda_p^2\lambda_{p+q}^2+ \lambda_q^2\lambda_{p+q}^2\Bigr) 
+4 (p+q)^2(\lambda_p^2+\lambda_q^2)
(\lambda_p\cdot\lambda_q)\right)  \biggr](s,t) \,,  \nonumber \\
A_{1234}^{\mathrm{DT}} & = i\mathcal{T}\, \mathcal{I}_4^{\mathrm{DT}}
\biggl[\frac{(D_s-2)^2}{2}\left(4p\cdot q+2(p-q)\cdot(k_1+k_2)-s\right)\lambda_p^2\lambda_q^2 \notag \\
&\hspace{2.2cm}
 +8(D_s-2)\left(\lambda_p^2+\lambda_q^2\right)
 \left(\lambda_p\cdot\lambda_q\right)
\left(p^2+q^2-(p-q)\cdot(k_1+k_2)+s\right) \biggr](s,t)   \,.
\label{eq:PDT}
\end{align}
The double-triangle integral 
displayed in \fig{Fig:PandNP}(c) is simply
\begin{align}
&\mathcal{I}_4^{\mathrm{DT}}[\mathcal{P}(\lambda_i,p,q,k_i)](s,t) =\frac{1}{s}
\,\mathcal{I}_4^{\mathrm{bow\hbox{-}tie}}[\mathcal{P}(\lambda_i,p,q,k_i)](s,t) 
\,,
\end{align}
so that all integrals in the new representation of the amplitude are given by trivalent graphs.
%

\begin{figure}[tb]
\def\hs{\hspace{.6cm}}
\def\scale{.35}
\centering
\subfigure[]{\includegraphics[scale=\scale]{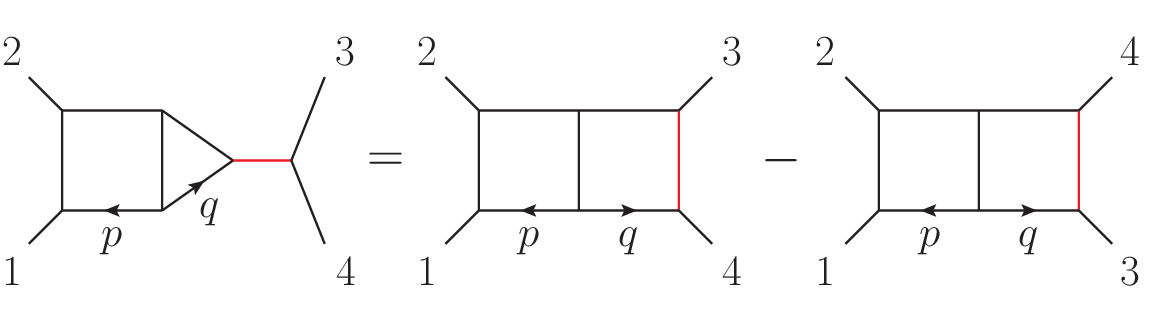}}
\hs
\subfigure[]{\includegraphics[scale=\scale]{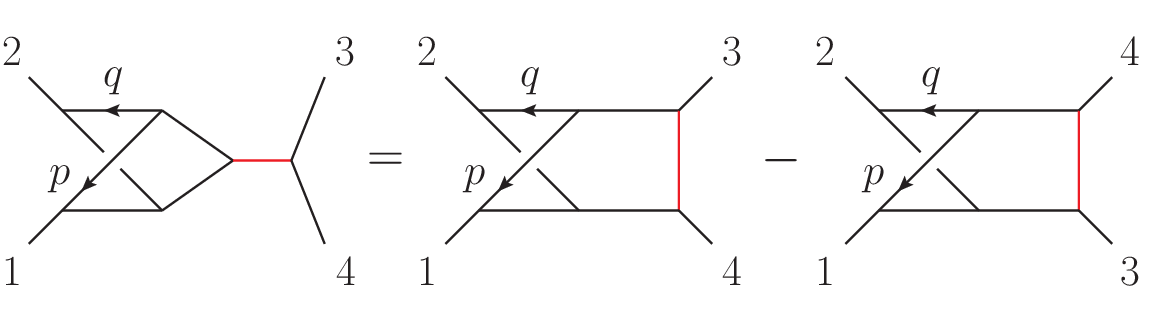}}
\subfigure[]{\includegraphics[scale=\scale]{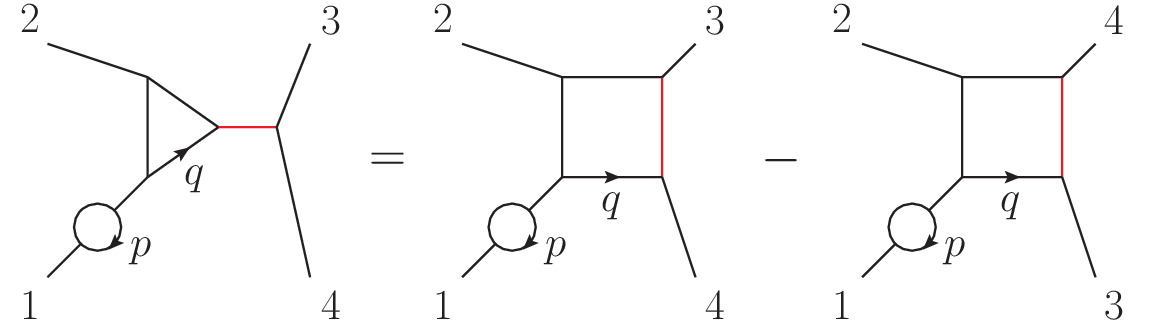}}
\hs
\subfigure[]{\includegraphics[scale=\scale]{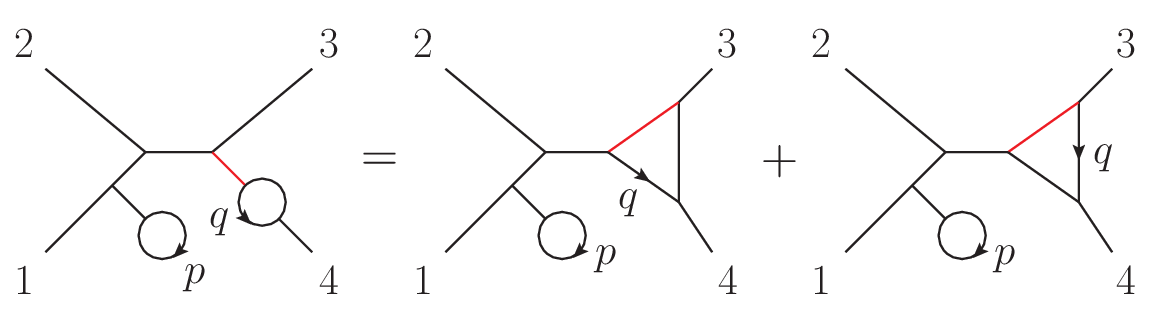}}
\subfigure[]{\includegraphics[scale=\scale]{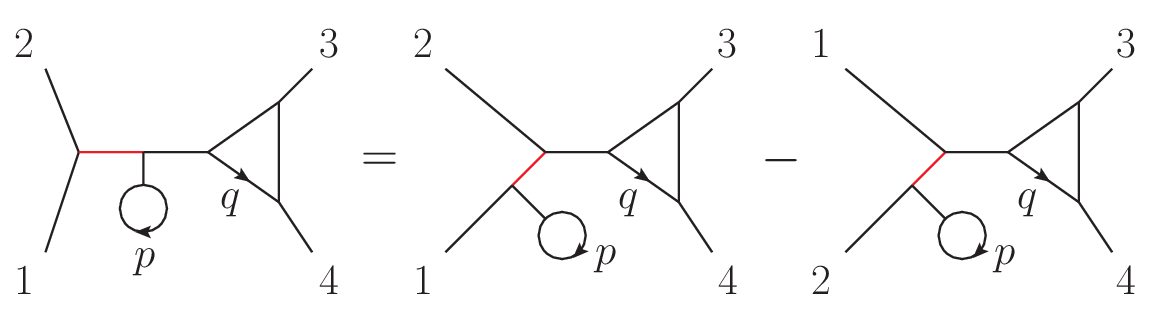}}
\hs
\subfigure[]{\includegraphics[scale=\scale]{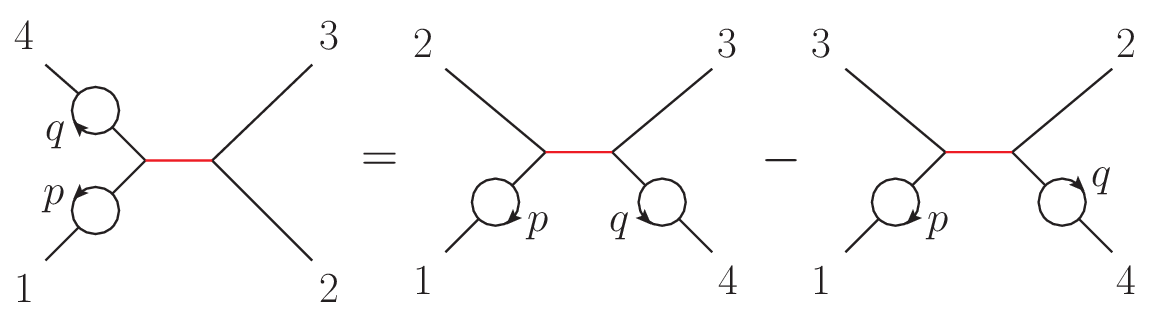}}
\caption[a]{Sample duality relations (a)-(f) involving graphs with vanishing
  numerators.  In each relation, the leftmost diagram has a vanishing
  numerator.  
  The shaded (red) leg marks the central leg of the applied
  dual-Jacobi identity.  }
\label{Fig:vanishDualityDiags}
\end{figure}

This form of the amplitude differs from Eq.~\eqref{eq:twoLoopYMOrig}
by absorbing the bow-tie contribution depicted in
\fig{Fig:bowtie} into both the planar double box in \fig{Fig:PandNP}(a) and the
 double triangle in \fig{Fig:PandNP}(c).
When moving terms into the double box (a),
we must multiply by a factor of
$(p+q)^2$ in the numerator to cancel the central propagator, while in
the double triangle (c), we must multiply by a factor of $s$.  In 
this rearrangement we have also included terms that 
integrate to zero.  In particular, the second term in the
double-triangle contribution in \eqn{eq:PDT} proportional to $(\lambda_p \cdot
\lambda_q)$ integrates to zero and does not contribute to the
integrated amplitude.  We are therefore free to drop it.  We can also
modify the first term in the double-triangle integral into the form
appearing in Ref.~\cite{Millenium} by using 
the fact that the substitution,
\begin{equation}
(4 p\cdot q + 2 (p-q)\cdot (k_1 + k_2) - s) \rightarrow 
2 (p+q)^2 +s\,,
\end{equation}
does not alter the value of the integrated amplitude: All terms that
are proportional to $p^2$, $q^2$, $(p-k_1-k_2)^2$, and $(q-k_3-k_4)^2$
yield scale-free integrals that integrate to zero.  Finally, to see
the equivalence of the two representations, we note that the double
triangle (c) has a different color factor from that of the planar
double box (a).  However, we can convert the double-triangle (c) color
factor to the double-box (a) color factor via the color Jacobi
identity $C^{\mathrm{DT}}_{1234} = C^{\mathrm{P}}_{1234} -
C^{\mathrm{P}}_{2134}$.  This matches the color assignment used in
Ref.~\cite{Millenium}.  Although not manifest, the kinematic numerator
reflects the antisymmetry of the Jacobi relations so that the
additional terms picked up by $A^{\mathrm{P}}_{1234}$ and
$A^{\mathrm{P}}_{2134}$ are simply related by relabelings.  Thus, after integration our
representation in \eqn{eq:twoLoopYM} is equivalent to the one in
\eqn{eq:twoLoopYMOrig}, which comes from Ref.~\cite{Millenium}.

The integrand in \eqn{eq:twoLoopYM} satisfies BCJ duality once we include
additional contributions that integrate to zero.  To find the full
form, we consider Jacobi relations (\ref{BCJDuality}) around each
internal propagator of the planar double box, the nonplanar double
box, and the double triangle, as well as all resultant integrals that
arise from these Jacobi relations.  Duality relations where all three
numerators are nonvanishing are depicted in \fig{Fig:twoLoopBCJ}.  The
need for additional nonvanishing numerators depicted in \fig{Fig:PandNP}(d)-(m)
arises from these dual-Jacobi
relations.
Other sample Jacobi relations where one of the numerators 
vanishes are shown in \fig{Fig:vanishDualityDiags}.
Up to relabelings, there are in total 16 such relations
involving two nonvanishing numerators and one vanishing numerator.
A fully duality-satisfying form is given by the numerators,
\begin{align}
&\mathcal{P}^{\mathrm{P}}(\lambda_i,p,q,k_i)=\frac{(D_s-2)^2}{2}(p+q)^2
  \lambda_p^2\lambda_q^2+16s\left((\lambda_p\cdot\lambda_q)^2
  -\lambda_p^2\lambda_q^2\right) \notag \\
&\hspace{3cm}+(D_s-2)\left(s\left(\lambda_p^2\lambda_q^2+
 \lambda_p^2\lambda_{p+q}^2+\lambda_q^2\lambda_{p+q}^2\right)
 +4(p+q)^2\left(\lambda_p^2+\lambda_q^2\right)
  \left(\lambda_p\cdot\lambda_q\right)\right)\,, \notag \\
&\mathcal{P}^{\mathrm{NP}}(\lambda_i,p,q,k_i)=(D_s-2)s
 \left(\lambda_p^2\lambda_q^2+\lambda_p^2\lambda_{p+q}^2
 +\lambda_q^2\lambda_{p+q}^2\right)
 +16s\left((\lambda_p\cdot\lambda_q)^2
 -\lambda_p^2\lambda_q^2\right)\,, \notag \\
&\mathcal{P}^{\mathrm{DT}}(\lambda_i,p,q,k_i)=
 \frac{(D_s-2)^2}{2}\left(4p\cdot q+2(p-q)\cdot(k_1+k_2)-s\right)
  \lambda_p^2\lambda_q^2 \notag \\
&\hspace{4.2cm}
 +8(D_s-2)\left(\lambda_p^2+\lambda_q^2\right)
  \left(\lambda_p\cdot\lambda_q\right)
  \left(p^2+q^2-(p-q)\cdot(k_1+k_2)+s\right)
   \vphantom{\frac{(D_s-2)^2}{2}}\,, \notag \\
&\mathcal{P}^{\mathrm{(d)}}(\lambda_i,p,q,k_i)=
  \frac{(D_s-2)^2}{2}(p+q)^2\lambda_p^2\lambda_q^2
  +4(D_s-2)(p+q)^2\left(\lambda_p^2+\lambda_q^2\right)
  \left(\lambda_p\cdot\lambda_q\right)\,, \notag \\
&\mathcal{P}^{\mathrm{(e)}}(\lambda_i,p,q,k_i)=(D_s-2)^2
 \left(p^2+q^2-(p-q)\cdot k_1\right)\lambda_p^2\lambda_q^2 \notag \\
&\hspace{4.6cm}+8(D_s-2)\left(2p\cdot q+(p-q)\cdot k_1\right)
 \left(\lambda_p^2+\lambda_q^2\right)\left(\lambda_p\cdot\lambda_q\right)  \notag \\
&\mathcal{P}^{\mathrm{(f)}}(\lambda_i,p,q,k_i) =
   -2(D_s-2)^2(p\cdot k_1)\lambda_p^2\lambda_q^2-16(D_s-2)
 (q\cdot k_1)\left(\lambda_p^2+\lambda_q^2\right)
  \left(\lambda_p\cdot\lambda_q\right) \notag \\
&\mathcal{P}^{\mathrm{(g)}}(\lambda_i,p,q,k_i)=
  \frac{(D_s-2)^2}{2}\left((p+q)^2\lambda_p^2+p^2\lambda_{p+q}^2\right)
  \lambda_q^2 \notag \\
&\hspace{3.4cm}+4(D_s-2)\left((p+q)^2\left(\lambda_p^2+\lambda_q^2\right)
 \left(\lambda_p\cdot\lambda_q\right)-p^2\left(\lambda_q^2+\lambda_{p+q}^2\right)
 \left(\lambda_q\cdot\lambda_{p+q}\right)\right)\,,  \notag \\
&\mathcal{P}^{\mathrm{(h)}}(\lambda_i,p,q,k_i)=2(D_s-2)^2
 \left((p\cdot q)\lambda_p^2+p^2\left(\lambda_p\cdot\lambda_q\right)\right)
  \lambda_q^2 \notag \\
&\hspace{4.6cm}-8(D_s-2)\left(3p^2\lambda_q^2-q^2
 \left(\lambda_p^2+\lambda_q^2\right)\right)
 \left(\lambda_p\cdot\lambda_q\right)\,, \notag \\
&\mathcal{P}^{\mathrm{(i)}}(\lambda_i,p,q,k_i)=-\frac{(D_s-2)^2}{2}(4q\cdot k_2+s)\lambda_p^2\lambda_q^2-4(D_s-2)(4p\cdot k_2-s)\left(\lambda_p^2+\lambda_q^2\right)\left(\lambda_p\cdot\lambda_q\right)\,, \notag \\
&\mathcal{P}^{\mathrm{(j)}}(\lambda_i,p,q,k_i)=8(D_s-2)s
\left(\lambda_p^2+\lambda_q^2\right)\left(\lambda_p\cdot\lambda_q\right)\,, \notag \\
&\mathcal{P}^{\mathrm{(k)}}(\lambda_i,p,q,k_i)=(D_s-2)^2t\,\lambda_p^2\lambda_q^2\,,
\label{IntegralNumerators}
\end{align}
where each $\mathcal{P}^{x}$ is the numerator of an integral
$\mathcal{I}_4^{x}[\mathcal{P}^{x}(\lambda_i,p,q,k_i)](s,t)$ 
corresponding to diagram $x$, depicted in \fig{Fig:PandNP}.
In contrast to the one-loop case, the duality-satisfying 
amplitudes do contain tadpole diagrams with nonvanishing numerators.

Although BCJ duality gives us a set of well-defined numerators for all
diagrams, the diagrams with on-shell or vanishing intermediate
propagators are ill-defined.  However, all such ill-defined diagrams
do not contribute to the standard two- and three-particle cuts and
 give vanishing contributions after integration.  In
more detail, ill-defined diagrams (e), (f), and
(h)--(k) in \fig{Fig:PandNP} contain scale-free integrals
that integrate to zero.  We also note that, using the numerators
in \eqn{IntegralNumerators}, well-defined diagrams (d) and (g) also
contain scale-free integrals
that vanish after integration.  Diagrams (f), (h), and (j) in
\fig{Fig:PandNP} contain a tadpole subdiagram.  We set these to zero,
just as they are set to zero in Feynman diagrams since the tadpole
integral is scale free in dimensional regularization.  Diagrams (e),
(i), and (k) are also ill-defined for on-shell external legs because
of the propagator carrying an on-shell momentum.  With Feynman
diagrams, this is normally dealt with by taking the legs off shell; in
principle, we can also define an off-shell continuation, although it
is nontrivial to do so consistently in our case.  However, such
ill-defined bubble-on-external-leg contributions again vanish in
dimensional regularization, since the integrals are also scale free.
In the gauge-theory case, although vanishing, these integrals can
potentially contain ultraviolet divergences that cancel completely
against infrared divergences. However, in the gravity case, which we
are interested in here, the integrals are suppressed by an additional
power of the on-shell invariant $k_i^2=0$ and therefore lead to
ultraviolet divergences with zero coefficient.  Diagrams (d) and (g)
in \fig{Fig:PandNP} may appear to have nonvanishing cut contributions,
but inverse propagators in the numerator cancel propagators, again
leaving scale-free integrals that vanish.

In summary, the two-loop four-point all-plus-helicity pure Yang-Mills
amplitude in a duality-satisfying representation is given by
\begin{align}
\mathcal{A}_4^{(2)}(1^+,2^+,3^+,4^+)&=g^6\sum_{\mathcal{S}_4}
   \sum_{x\in\{\mathrm{diagrams}\}}\frac{1}{S^x}C_{1234}^xA_{1234}^x\,,
\label{eq:twoLoopYMBCJ} 
\end{align}
where $x$ labels diagrams in \fig{Fig:PandNP}
with nonvanishing numerators.  $S^x$ is
the symmetry factor of diagram $x$, while $C_{1234}^x$ is the color
factor.  The partial amplitudes are given by
\begin{align}
&A_{1234}^x=i\mathcal{T}\, \mathcal{I}_4^x[\mathcal{P}^x(\lambda_i,p,q,k_i)](s,t)\,,
\end{align}
where all diagrams except for those in \fig{Fig:PandNP}(a), (b), (c)
integrate to zero in gauge theory.  In Section~\ref{sec:twoLoopUV}, we
will use the double-copy relation (\ref{DoubleCopy}) on these
numerators to study the two-loop ultraviolet behavior of gravity coupled to a
dilaton and an antisymmetric tensor.

\section{Ultraviolet Properties of Gravity}
\label{sec:LoopUV}

We now turn to the ultraviolet properties of the gravity double-copy
theory consisting of a graviton, dilaton, and antisymmetric tensor,
from the perspective of the double-copy formalism.  The theory
generated by taking the double copy of pure Yang-Mills corresponds to
the low-energy effective Lagrangian of the bosonic part of string
theory~\cite{EffectiveAction},
\begin{align}
\mathcal{L}=\sqrt{-g}\left(\frac{2}{\kappa^2}R
+\frac{1}{2}\partial_{\mu}\phi\partial^{\mu}\phi
+\frac{1}{6} e^{-2\kappa \phi/\sqrt{D-2}} H_{\mu\nu\rho}H^{\mu\nu\rho}\right),
\label{eq:Lagrangian}
\end{align}
where
$H_{\mu\nu\rho}=\partial_{\mu}A_{\nu\rho}+\partial_{\nu}A_{\rho\mu}+\partial_{\rho}A_{\mu\nu}$,
and $A_{\mu\nu}=-A_{\nu\mu}$ is the rank-two antisymmetric tensor
field.

Pure Einstein gravity is one-loop finite in four
dimensions~\cite{tHooftVeltman}.  However, when coupled to a scalar
(dilaton)~\cite{tHooftVeltman} or to a rank-two antisymmetric
tensor~\cite{antisymm}, the theory is divergent.  We find that the
double-copy theory coupled to both a dilaton and an antisymmetric
tensor is also divergent, although for all these theories the
four-point amplitudes with at least one external graviton are finite,
as expected from simple counterterm arguments.  We will show that the
cancellation no longer holds at two loops, and the theory has an $R^3$
counterterm, in much the same way as it does for pure Einstein
gravity~\cite{GoroffSagnotti}.  In six dimensions, pure Einstein
gravity is ultraviolet divergent at one loop~\cite{Deq6Div}.  We find
the same to be true in our double-copy theory, and we find a
divergence in eight dimensions as well.  We will give the explicit
form of the divergences for these cases.  In carrying out these
computations we use the four-dimensional helicity scheme~\cite{FDH}.
It would be interesting to compare our results to ones obtained using
the standard dimensional-regularization scheme, used in, for example,
Ref.~\cite{GoroffSagnotti}.

\subsection{One Loop}

\subsubsection{Four Dimensions}
\label{sec:OneLoop4D}

In four dimensions, there is no one-loop four-point divergence when
one external leg is a graviton~\cite{tHooftVeltman, antisymm} because
the potential independent counterterms for such divergences vanish on
shell or can be eliminated by the equations of motion.  Using the
double-copy formula (\ref{DoubleCopy}), we have explicitly confirmed
finiteness in one-loop four-point amplitudes containing at least one
external graviton, with the remaining legs either gravitons, dilatons, or
antisymmetric tensors.  We obtain the gravity numerator from the
double-copy formula (\ref{DoubleCopy}) by taking the two Yang-Mills
numerators, $\tilde n_i$ and $n_i$, to be equal to the BCJ form of the
Yang-Mills numerator (\ref{YMNum}).  As an interesting cross check, we
have obtained an asymmetric representation of the gravity amplitudes
by taking the $\tilde n_i$ to be the numerators that satisfy BCJ
duality and the $n_i$ to be numerators obtained by gauge-theory
Feynman rules in Feynman gauge, similar to the procedure used recently
for half-maximal supergravity~\cite{ThreeloopHalfMax,TwoloopHalfMax}.
By generalized gauge invariance~\cite{BCJLoop,YMSquared}, this should
be equivalent to the symmetric construction.  Indeed, we find
identical results for the ultraviolet divergences.

To evaluate the ultraviolet divergences, we expand in small external
momenta to reduce to logarithmically divergent
integrals~\cite{smallMomenta}.  We then simplify tensor integrals
composed of loop momenta in the numerators by using Lorentz
invariance, which implies that the integrals must be linear combinations of
products of metric tensors $\eta^{\mu\nu}$.  (See Ref.~\cite{ck4l} for
a recent discussion of evaluating tensor vacuum integrals.)  With the
insertion of a massive infrared regulator, we finally integrate simple
one-loop integrals to find the potential ultraviolet divergence.  Due
to our auxiliary conditions, contributions from bubbles on external legs
vanish, as they would for ordinary gravity Feynman diagrams. We
therefore obtain our entire result from box, triangle, and
bubble-on-internal-leg diagrams.

For completeness we have also computed the divergences directly in
dimensional regularization without introducing a mass regulator, using
techniques similar to those for two loops in
Appendix~\ref{sec:DimRegAppendix}.  After subtracting the infrared divergence
as computed in Appendix~\ref{sec:IRAppendix}, we find complete agreement with
our result found using vacuum integrals.

We obtain an expression for the divergence in terms of formal
polarization vectors.  By taking linear combinations of the product of
polarization vectors from each copy of Yang-Mills, we can project onto
the graviton, dilaton, and antisymmetric tensor states.  In $D=4$ this
is conveniently implemented by using spinor
helicity~\cite{SpinorHelicity}.  Graviton polarization tensors
correspond to the `left' and `right' copies of Yang-Mills according to
$\varepsilon_{\mu\nu}^{h+}\rightarrow\varepsilon_{\mathrm{L}\mu}^+\varepsilon_{\mathrm{R}\nu}^+$
and
$\varepsilon_{\mu\nu}^{h-}\rightarrow\varepsilon_{\mathrm{L}\mu}^-\varepsilon_{\mathrm{R}\nu}^-$.
For the dilaton and antisymmetric tensor, we symmetrize and
antisymmetrize in opposite-helicity configurations according to
$\varepsilon_{\mu\nu}^{\phi}\rightarrow\frac{1}{\sqrt{2}}
(\varepsilon_{\mathrm{L}\mu}^+\varepsilon_{\mathrm{R}\nu}^-
+\varepsilon_{\mathrm{L}\mu}^-\varepsilon_{\mathrm{R}\nu}^+)$
and
$\varepsilon_{\mu\nu}^{A}\rightarrow\frac{1}{\sqrt{2}}
(\varepsilon_{\mathrm{L}\mu}^+\varepsilon_{\mathrm{R}\nu}^-
-\varepsilon_{\mathrm{L}\mu}^-\varepsilon_{\mathrm{R}\nu}^+)$.
By substituting the explicit polarizations, we find that all
configurations where at least a single leg is a graviton are free of
ultraviolet divergences,
\begin{equation}
\mathcal{M}^{(1)}(1^h, 2, 3, 4) \Bigr|_{\rm div.} = 0\,,
\end{equation}
where leg $1$ is either a positive- or negative-helicity graviton, and
the other three states are unspecified.

We however find divergences for the cases with no external gravitons.
For the four-dilaton amplitude, we find 
\begin{equation}
\mathcal{M}^{(1)}(1^{\phi},2^{\phi},3^{\phi},4^{\phi}) \Bigr|_{\rm div.}
=\frac{1}{\epsilon}\left(\frac{\kappa}{2}\right)^4
\frac{i}{(4\pi)^2}\frac{1132-92D_s+3D_s^2}{120}\left(s^2+t^2+u^2\right),
\label{eq:scalarDiv}
\end{equation}
corresponding to the operator,
\begin{equation}
\frac{1}{\epsilon}\left(\frac{\kappa}{2}\right)^4
\frac{1}{(4\pi)^4}\frac{1132-92D_s+3D_s^2}{240}
(D_{\mu}\phi D^{\mu}\phi)^2\,.
\end{equation}
This result is similar to the one obtained long ago by 't~Hooft and
Veltman~\cite{tHooftVeltman}.  However, in our case we have an
antisymmetric tensor which can circulate in the loop, altering the
numerical coefficient.  We note that the operator in
Ref.~\cite{tHooftVeltman} looks different than above, but it can be
written in a similar way through use of the field equations of motion.

The amplitude with four antisymmetric tensors is also one-loop divergent in four
dimensions. In four dimensions, the antisymmetric tensor is
dual to a scalar field, so we expect the divergence to be the
same as that for dilatons.  Indeed, the divergence in the four-antisymmetric-tensor 
amplitude for a theory with an antisymmetric 
tensor coupled to gravity is equal to that of the four-dilaton amplitude
in a theory of a dilaton coupled to gravity~\cite{antisymm}.  In
congruence, we find the divergence for four external antisymmetric
tensors to also be given by the same expression as the four-dilaton 
divergence \eqref{eq:scalarDiv},
\begin{equation}
\mathcal{M}^{(1)} (1^A, 2^A, 3^A, 4^A) \Bigr|_{\rm div.} = 
\mathcal{M}^{(1)}(1^{\phi},2^{\phi},3^{\phi},4^{\phi}) \Bigr|_{\rm div.}\, .
\end{equation}
In terms of the
antisymmetric tensor fields, the divergence is generated by the operator,
\begin{equation}
\frac{1}{\epsilon}\left(\frac{\kappa}{2}\right)^4
\frac{1}{(4\pi)^4}\frac{1132-92D_s+3D_s^2}{2160}\left(H_{\mu\nu\rho}H^{\mu\nu\rho}\right)^2.
\end{equation}
The counterterm that cancels the divergence is given by the negative
of this operator.

In addition to the above divergences, there is also a divergence in the
$D=4$, $\phi \phi AA$ amplitude.  This divergence is
given by
\begin{eqnarray}
\mathcal{M}^{(1)}(1^{\phi},2^{\phi},3^{A},4^{A})\Bigr|_{\rm div.} &= &
\frac{1}{\epsilon}\left(\frac{\kappa}{2}\right)^4\frac{i}{(4\pi)^4}
\biggr(\frac{1116-76D_s-D_s^2}{120}\,s^2 \nonumber \\
&&\hspace{3cm} \null 
+\frac{-1124+84D_s-D_s^2}{120}\,(t^2+u^2) \biggr),  \hskip .5 cm 
\end{eqnarray}
which corresponds to the operator, 
\begin{eqnarray}
&& \frac{1}{\epsilon}\left(\frac{\kappa}{2}\right)^4\frac{1}{(4\pi)^4}
\biggl(\frac{1124-84D_s+D_s^2}{60}\,
   H_{\mu\rho\sigma}H_{\nu}{}^{\rho\sigma}D^{\mu}\phi D^{\nu}\phi \nonumber \\
&&\hspace{2.8cm} \null 
 - \frac{1132-92D_s+3D_s^2}{360}\,
    H_{\mu\nu\rho}H^{\mu\nu\rho} D_{\sigma}\phi D^{\sigma}\phi\biggr) \,.
\end{eqnarray}

\subsubsection{Six Dimensions}

In six dimensions for external gravitons, the only independent
invariant operator at one loop~\cite{RCubed} is
\begin{equation}
R_{\alpha\beta\mu\nu}R^{\mu\nu\rho\sigma}R_{\rho\sigma}{}^{\alpha\beta}.
\label{eq:RCubed}
\end{equation}
This corresponds to the known $D=6$ one-loop divergence 
of pure Einstein gravity given in 
Ref.~\cite{Deq6Div}.   We have computed the coefficient of the 
$D=6$ divergence for the double-copy theory of a graviton
coupled to a dilaton and an
antisymmetric tensor. In this case, the divergence is given by the operator,
\begin{equation}
-\frac{1}{\epsilon}\frac{1}{(4\pi)^3}\frac{(D_s-2)^2}{30240}R_{\alpha\beta\mu\nu}R^{\mu\nu\rho\sigma}R_{\rho\sigma}{}^{\alpha\beta}.
\end{equation}
Appropriate powers of the coupling are generated by expanding the
metric around flat space, $g_{\mu\nu}=\eta_{\mu\nu}+\kappa
h_{\mu\nu}$.  Although we do not include the explicit forms of the
counterterms here, we have also found divergences for the following
amplitudes (as well as their permutations and parity conjugates) involving
external dilatons and antisymmetric tensors, where we restrict the
external states to four dimensions:
\begin{eqnarray}
&& \mathcal{M}^{(1)}(1^{\phi},2^+,3^+,4^+)\,,  \hskip .2 cm 
\mathcal{M}^{(1)}(1^{\phi},2^{\phi},3^+,4^+)\,, \hskip .2 cm 
\mathcal{M}^{(1)}(1^{\phi},2^{\phi},3^{\phi},4^+)\,, \hskip .2 cm 
\mathcal{M}^{(1)}(1^{\phi},2^{\phi},3^{\phi},4^{\phi})\,, \hskip 1 cm \nonumber \\
&& \mathcal{M}^{(1)}(1^A,2^A,3^+,4^+)\,,   \hskip .2 cm 
\mathcal{M}^{(1)}(1^A,2^A,3^{\phi},4^+) \,,  \hskip .2 cm 
\mathcal{M}^{(1)}(1^A,2^A,3^{\phi},4^{\phi}) \,.
\end{eqnarray}

\subsubsection{Eight Dimensions}

In eight dimensions, there are seven linearly independent $R^4$
operators \cite{RFourth}:
\begin{eqnarray}
T_1&=&(R_{\mu\nu\rho\sigma}R^{\mu\nu\rho\sigma})^2\,, \nonumber \\
T_2&=&R_{\mu\nu\rho\sigma}R^{\mu\nu\rho}_{\hphantom{\mu\nu\rho}\lambda}R_{\gamma\delta\kappa}^{\hphantom{\gamma\delta\kappa}\sigma}R^{\gamma\delta\kappa\lambda}\,, \nonumber \\
T_3&=&R_{\mu\nu\rho\sigma}R^{\mu\nu}_{\hphantom{\mu\nu}\lambda\gamma}R^{\lambda\gamma}_{\hphantom{\lambda\gamma}\delta\kappa}R^{\rho\sigma\delta\kappa}\,, \nonumber \\
T_4&=&R_{\mu\nu\rho\sigma}R^{\mu\nu}_{\hphantom{\mu\nu}\lambda\gamma}R^{\rho\lambda}_{\hphantom{\rho\lambda}\delta\kappa}R^{\sigma\gamma\delta\kappa}\,, \nonumber \\
T_5&=&R_{\mu\nu\rho\sigma}R^{\mu\nu}_{\hphantom{\mu\nu}\lambda\gamma}R^{\rho\hphantom{\delta}\lambda}_{\hphantom{\rho}\delta\hphantom{\lambda}\kappa}R^{\sigma\delta\gamma\kappa}\,, \nonumber \\
T_6&=&R_{\mu\nu\rho\sigma}R^{\mu\hphantom{\lambda}\rho}_{\hphantom{\mu}\lambda\hphantom{\rho}\gamma}R^{\lambda\hphantom{\delta}\gamma}_{\hphantom{\lambda}\delta\hphantom{\gamma}\kappa}R^{\nu\delta\sigma\kappa}\,, \nonumber \\
T_7&=&R_{\mu\nu\rho\sigma}R^{\mu\hphantom{\lambda}\rho}_{\hphantom{\mu}\lambda\hphantom{\rho}\gamma}R^{\lambda\hphantom{\delta}\nu}_{\hphantom{\lambda}\delta\hphantom{\nu}\kappa}R^{\gamma\delta\sigma\kappa} \,.
\end{eqnarray}
On shell, the combination,
\begin{equation}
U=-\frac{T_1}{16}+T_2-\frac{T_3}{8}-T_4+2T_5-T_6+2T_7\,,
\label{eq:U}
\end{equation}
is a total derivative, so only six of the $T_i$ are independent on
shell.  In terms of these operators, the divergence for gravity
coupled to a dilaton and an antisymmetric tensor at one loop in $D=8$
is 
\begin{eqnarray}
&& \frac{1}{\epsilon}\frac{1}{(4\pi)^4}\frac{1}{1814400}
[(4274-899D_s+11D_s^2)T_1-40(466-103D_s-2D_s^2)T_2 \nonumber \\
&& \hskip 3 cm \null -2(1886+319D_s-D_s^2)T_3-180(1034+D_s)T_4  \\
&& \hskip 3 cm \null + 16(1196+34D_s-D_s^2)T_6+64(12454+71D_s+D_s^2)T_7+c\,U]\,,
\hskip 1.5 cm \vphantom{\frac{1}{1814400}} \nonumber
\end{eqnarray}
where $c$ is a free parameter multiplying the total derivative~\eqref{eq:U} .

We have also found that the following four-point amplitudes involving
dilatons and antisymmetric tensors diverge in $D=8$:
\begin{eqnarray}
&&\mathcal{M}^{(1)}(1^{\phi},2^{\phi},3^+,4^+)\,, \hskip .3 cm 
\mathcal{M}^{(1)}(1^{\phi},2^{\phi},3^+,4^-)\,, \hskip .25 cm 
\mathcal{M}^{(1)}(1^{\phi},2^{\phi},3^{\phi},4^{\phi})\,, \hskip .25 cm 
\mathcal{M}^{(1)}(1^A,2^A,3^+,4^+)\,, \nonumber \\
&& \mathcal{M}^{(1)}(1^A,2^A,3^+,4^-)\,, \hskip .2 cm 
\mathcal{M}^{(1)}(1^A,2^A,3^{\phi},4^+)\,, \hskip .2 cm  
\mathcal{M}^{(1)}(1^A,2^A,3^{\phi},4^{\phi})\,, \hskip .2 cm 
\mathcal{M}^{(1)}(1^A,2^A,3^A,4^A) \,,\nonumber   \\
\end{eqnarray}
where we have again chosen the external states to be four
dimensional.  The other configurations are finite.

\subsection{Ultraviolet Properties of Gravity at Two Loops in Four Dimensions}
\label{sec:twoLoopUV}

Pure Einstein gravity in $D=4$ is one-loop finite, but it does diverge
at two loops~\cite{GoroffSagnotti}.  This suggests that the two-loop
four-graviton amplitude, including also the dilaton and antisymmetric
tensor, should diverge as well.  For external gravitons, the only
independent operator is the same $R^3$ operator for one loop in six
dimensions \eqref{eq:RCubed}.  Our aim is to find its
coefficient.

\begin{figure}
\includegraphics[scale=.4]{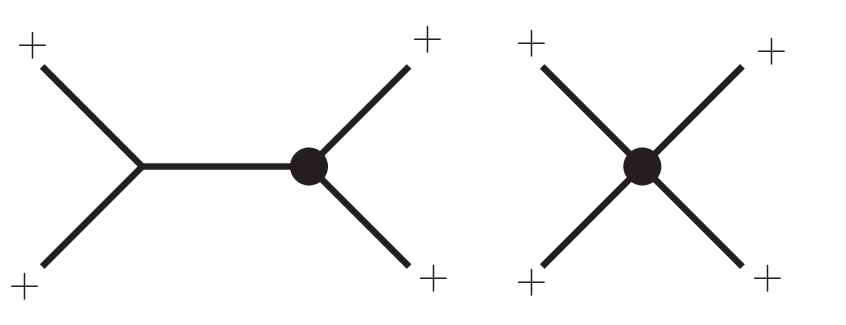}
\caption{The $R^3$ operator diagrams that contribute to the all-plus-helicity four-graviton amplitude. The solid dot represents vertices generated
by the $R^3$ operator. }
\label{fig:Counterterm}
\end{figure}

The $R^3$ operator generates a nonvanishing four-point amplitude for
identical-helicity gravitons, illustrated in \fig{fig:Counterterm}.
This means that we can determine the coefficient of this operator by
computing the four-graviton all-plus-helicity amplitude.  Fortunately,
as we discussed in \sect{sec:Construction}, we have the BCJ form of
the required all-plus-helicity Yang-Mills amplitude.  Applying the
double-copy formula (\ref{DoubleCopy}) to the Yang-Mills amplitude in
\eqn{eq:twoLoopYMBCJ} immediately gives us the corresponding gravity
integrand, simply by squaring the numerators.  Diagrams (d)-(k) in
Fig.~\ref{Fig:PandNP} integrate to zero in gravity just as they did in
Yang-Mills.  In addition, as was mentioned in
Section~\ref{sec:twoLoopBCJ}, the second term of the double-triangle
in Eq.~\eqref{eq:PDT} also integrates to zero; in fact, 
due to the simple identity,
\begin{equation}
-(p-q)\cdot (k_1 + k_2) +s = \frac{1}{2}(p-k_1-k_2)^2 + \frac{1}{2} (q+k_1+k_2)^2 -
                             \frac{1}{2} p^2 -  \frac{1}{2} q^2 \, ,
\end{equation}
all such terms will integrate to zero because the inverse propagators lead to 
scale-free integrals.  Thus, the four-graviton all-plus-helicity amplitude is
given by
\begin{equation}
\mathcal{M}^{(2)}(1^+,2^+,3^+,4^+)
= \left(\frac{\kappa}{2}\right)^6
\sum_{\mathcal{S}_4}\left[\frac{1}{4}M_{1234}^{\mathrm{P}}+\frac{1}{4}M_{12;34}^{\mathrm{NP}}+\frac{1}{8}M_{1234}^{\mathrm{DT}}\right]\,,
\label{eq:twoLoopGrav}
\end{equation}
where
\begin{align}
&M_{1234}^{\mathrm{P}}=i \, \mathcal{T}^2\,
\mathcal{I}_4^{\mathrm{P}}\biggl[\biggl(
\frac{(D_s-2)^2}{2}(p+q)^2\lambda_p^2\lambda_q^2
+ 16s\left((\lambda_p\cdot\lambda_q)^2-\lambda_p^2\lambda_q^2\right) \nonumber \\
& \hskip 2.1 cm \null
+(D_s-2) \Bigl(s\left(\lambda_p^2\lambda_q^2+\lambda_p^2\lambda_{p+q}^2+ \lambda_q^2\lambda_{p+q}^2\Bigr) 
+4 (p+q)^2(\lambda_p^2+\lambda_q^2)
(\lambda_p\cdot\lambda_q)\right) \biggr)^{\!\!2} \biggr](s,t) \nonumber \\
&\hphantom{M_{1234}^{\mathrm{P}}}=i \, \mathcal{T}^2
\Bigg\{\mathcal{I}_4^{\mathrm{P}}\Bigl[
\Bigl((D_s-2)s\left(\lambda_p^2\lambda_q^2+\lambda_p^2\lambda_{p+q}^2+
\lambda_q^2\lambda_{p+q}^2\right)+16s\left((\lambda_p\cdot\lambda_q)^2-\lambda_p^2\lambda_q^2\right)\Bigr)^{\!2}\,\Bigr](s,t) \nonumber \\
&\hspace{1.5cm}\null
+\mathcal{I}_4^{\mathrm{bow\hbox{-}tie}}\left[2\left(4(D_s-2)(\lambda_p^2+\lambda_q^2)(\lambda_p\cdot\lambda_q)+\frac{(D_s-2)^2}{2}\lambda_p^2\lambda_q^2\right)\right. \nonumber \\
&\hspace{3.2cm}\left.\vphantom{\frac{(D_s-2)^2}{s}}\times\left((D_s-2)s\left(\lambda_p^2\lambda_q^2+\lambda_p^2\lambda_{p+q}^2+
\lambda_q^2\lambda_{p+q}^2\right)+16s\left((\lambda_p\cdot\lambda_q)^2-\lambda_p^2\lambda_q^2\right)\right)\right]\!(s,t) \nonumber \\
&\hspace{1.5cm}\null 
+\mathcal{I}_4^{\mathrm{bow\hbox{-}tie}}\biggl[
(p+q)^2\biggl(\frac{(D_s-2)^2}{2}\lambda_p^2\lambda_q^2+4(D_s-2)(\lambda_p^2+\lambda_q^2)(\lambda_p\cdot\lambda_q)\biggr)^{\!\! 2}\,\biggr](s,t)\Bigg\}\,, \nonumber  \\
&M_{12;34}^{\mathrm{NP}}=i\, \mathcal{T}^2s^2 \,
\mathcal{I}_4^{\mathrm{NP}}\Bigl[
\Bigl((D_s-2)\left(\lambda_p^2\lambda_q^2+\lambda_p^2\lambda_{p+q}^2+
\lambda_q^2\lambda_{p+q}^2\right)+16\left((\lambda_p\cdot\lambda_q)^2-\lambda_p^2\lambda_q^2\right)\Bigr)^{\!2}\,\Bigr](s,t)\,, \nonumber \\
&M_{1234}^{\mathrm{DT}}=i\, \mathcal{T}^2\,\mathcal{I}_4^{\mathrm{DT}}\biggl[\biggl(\frac{(D_s-2)^2}{2}\left(4p\cdot q+2(p-q)\cdot(k_1+k_2)-s\right)\lambda_p^2\lambda_q^2\biggr)^{\!2}\,\biggr](s,t) \nonumber \\
&\hphantom{M_{1234}^{\mathrm{DT}}}=i\, \mathcal{T}^2\frac{1}{s}\mathcal{I}_4^{\mathrm{bow\hbox{-}tie}}\biggl[\biggl(\frac{(D_s-2)^2}{2}\left(4p\cdot q+2(p-q)\cdot(k_1+k_2)-s\right)\lambda_p^2\lambda_q^2\biggr)^{\!2}\,\biggr](s,t)\,.
\label{eq:twoLoopDoubleCopy}
\end{align}
We have explicitly confirmed that $s$-, $t$-, and $u$-channel unitarity
cuts are satisfied.  We did so numerically keeping the internal 
states in integer dimensions $D=6$
and $D=8$.

To obtain the ultraviolet divergences, we integrate the amplitudes in
dimensional regularization.  We
carry out the extraction of the ultraviolet divergences in two
ways. In the first approach we simply use dimensional regularization
and then subtract the known infrared divergences, leaving only the
ultraviolet ones.  In the second approach we introduce a mass
regulator to separate the ultraviolet singularities from the infrared
divergences, as carried out in Appendix~\ref{sec:UVIntegralsAppendix}.
Either method yields the same result.  In fact, the second method also
shows that the vanishing integrals that we dropped, including diagrams (d)-(k) in Fig.~\ref{Fig:PandNP} and the second term of the double-triangle in Eq.~\eqref{eq:PDT}, are not ultraviolet divergent.

The dimensionally regularized integrals are performed in
Appendix~\ref{sec:DimRegAppendix}.\footnote{We thank L.~Dixon for
  cross-checking our integrals.}  Eq.~\eqref{eq:dimRegPlanar} gives
the planar double-box integrals; Eq.~\eqref{eq:dimRegNonplanar} gives
the nonplanar double-box integrals; and Eq.~\eqref{eq:bowTieInts}
gives the bow-tie integrals.  The infrared divergence from
Appendix~\ref{sec:IRAppendix} is
\begin{align}
\left.\mathcal{M}^{(2)}(1^+,2^+,3^+,4^+)\right|_{\mathrm{IR~div.}}=&-\frac{1}{\epsilon}\left(\frac{\kappa}{2}\right)^6\frac{i}{(4\pi)^4}\mathcal{T}^2\,\frac{(D_s-2)^2}{120}\left(s^2+t^2+u^2\right) \nonumber \\
&\hspace{.5cm}\times\left[s\,\mathrm{log}\left(\frac{-s}{\mu^2}\right)+t\,\mathrm{log}\left(\frac{-t}{\mu^2}\right)+u\,\mathrm{log}\left(\frac{-u}{\mu^2}\right)\right].
\end{align}
We insert the divergent parts of the integrals evaluated using
dimensional regularization into Eq.~\eqref{eq:twoLoopDoubleCopy}, then
insert these results into Eq.~\eqref{eq:twoLoopGrav} and perform the
permutation sum.  Finally we subtract the infrared divergence and arrive at
the two-loop ultraviolet divergence of gravity coupled to a dilaton and an
antisymmetric tensor for four external positive-helicity 
gravitons:
\begin{align}
\left.\mathcal{M}^{(2)}(1^+,2^+,3^+,4^+)\right|_{\mathrm{UV~div.}}&=\frac{1}{\epsilon}\left(\frac{\kappa}{2}\right)^6\frac{i}{(4\pi)^4}\mathcal{T}^2 \notag \\
&\hspace{1cm}\times\frac{(2D_s^4-136D_s^3+2883D_s^2-35164D_s+103052)stu}{10800}\,.
\label{eq:twoLoopUV}
\end{align}

For our second method, we evaluate the ultraviolet divergences of the required
integrals by going to vacuum and using a massive infrared regulator,
sidestepping the need to subtract the infrared divergence.  The ultraviolet
divergences of the individual integrals are calculated in
Appendix~\ref{sec:UVIntegralsAppendix}.  After permutations, the contributions
of the planar double-box, nonplanar double-box, and double-triangle components are
\begin{align}
&\left.\mathcal{M}^{\mathrm{P}}(1^+,2^+,3^+,4^+)\right|_{\mathrm{UV~div.}}
= -\frac{1}{\epsilon}\left(\frac{\kappa}{2}\right)^6\hspace{-.1cm}\frac{i}{(4\pi)^4}\mathcal{T}^2\,
\frac{(2D_s^3-63D_s^2+588D_s-1420)stu}{180}\,, \nonumber \\
&\left.\mathcal{M}^{\mathrm{NP}}(1^+,2^+,3^+,4^+)\right|_{\mathrm{UV~div.}}=-\frac{1}{\epsilon}\left(\frac{\kappa}{2}\right)^6\hspace{-.1cm}\frac{i}{(4\pi)^4}\mathcal{T}^2\,\frac{(21D_s^2-4D_s-396)s t u}{240}\,, \nonumber \\
&\left.\mathcal{M}^{\mathrm{DT}}(1^+,2^+,3^+,4^+)\right|_{\mathrm{UV~div.}}= \frac{1}{\epsilon}\left(\frac{\kappa}{2}\right)^6\hspace{-.1cm}\frac{i}{(4\pi)^4}\mathcal{T}^2\,\frac{(D_s-2)^4stu}{5400}\,.
\end{align}
Summing these contributions, we find complete agreement with Eq.~\eqref{eq:twoLoopUV}.

We can re-express the two-loop divergence in terms of the operator that
generates it.  By matching the amplitude generated by the 
diagrams with an $R^3$ vertex shown in \fig{fig:Counterterm} to the divergence 
in \eqn{eq:twoLoopUV}, we find that the operator, 
\begin{align}
-\frac{1}{\epsilon}\left(\frac{\kappa}{2}\right)^2\frac{1}{(4\pi)^4}
\frac{2D_s^4-136D_s^3+2883D_s^2-35164D_s+103052}{648000}\, R_{\alpha\beta\mu\nu}
    R^{\mu\nu\rho\sigma}R_{\rho\sigma}{}^{\alpha\beta},
\end{align}
generates the two-loop divergence for gravity coupled to a dilaton and
an antisymmetric tensor.

\section{Conclusion}
\label{sec:Conclusion}

In this paper we constructed a representation of the one-loop
four-point amplitude of pure Yang-Mills theory explicitly exhibiting
the duality between color and kinematics.  This
construction is the first nonsupersymmetric example at loop level
valid in any dimension with no restriction on the external states. 
The cost of this generality is relatively complicated 
expressions in terms of formal polarization vectors.

The duality between color and kinematics and its associated gravity
double-copy structure has proven useful for unraveling ultraviolet
properties in various dimensions~\cite{BCJLoop,ck4l,ThreeloopHalfMax,
  TwoloopHalfMax, BoelsUV}.  Using the one-loop four-point pure
Yang-Mills amplitude with the duality manifest, we obtained the
integrand for the corresponding amplitude in a theory of a graviton,
dilaton, and antisymmetric tensor.  In $D=4$, we found that one-loop
four-point amplitudes with one or more external gravitons are
ultraviolet finite, while amplitudes involving only external dilatons
or antisymmetric tensor fields diverge.  This result is similar to
those of earlier studies involving gravity coupled either to a scalar,
an antisymmetric tensor, or other matter and is in line with simple
counterterm arguments~\cite{tHooftVeltman, OtherGravityMatter,
  antisymm}.  We gave the explicit form, including numerical
coefficients, for all four-point divergences in this theory.  Since
our construction is valid in any dimension, we also investigated the
ultraviolet properties of the double-copy theory in higher dimensions.
In particular, we showed that in $D=6,8$ the one-loop four-graviton
amplitudes diverge, as expected, and gave the explicit form of these
divergences including their numerical coefficients.

In order to investigate whether the observed $D=4$ ultraviolet
finiteness of the amplitudes with one or more external gravitons
continues beyond one loop, we also computed the coefficient of the
potential two-loop $R^3$ divergences.  This was greatly simplified by
the observation that the coefficient of the divergence can be determined
from the
identical-helicity four-graviton configuration.  The required gravity
amplitude was then easily constructed via the double-copy property, by
first finding a representation of the pure Yang-Mills amplitude that
satisfies the duality. The existence of such a representation has
already been noted in Ref.~\cite{BCJLoop}.  Here we provided the
explicit representation, including diagrams that integrate to zero not
present in the original form of the two-loop
identical-helicity amplitude given in Ref.~\cite{Millenium}.  We found that
the two-loop amplitude with external gravitons is indeed divergent and
that the $R^3$ counterterm has nonzero coefficient.  This is not
surprising given that pure Einstein gravity diverges at two
loops~\cite{GoroffSagnotti}.  Our paper definitively shows that, as
one might have expected, the double-copy property by itself cannot
render a gravity theory ultraviolet finite.  For ultraviolet
finiteness, an additional mechanism such as supersymmetry is needed.
Further progress in clarifying the ultraviolet structure of gravity
theories will undoubtedly rely on new multiloop calculations to guide
theoretical developments.  We expect that the duality between color
and kinematics will continue to play an important role in this.

\subsection*{Acknowledgments}

We thank J.~J.~M.~Carrasco, H.-H.~Chi, L.~Dixon, H.~Elvang, H.~Johansson,
S.~Naculich, 
R.~Roiban, E.~Serna Campillo, and A.~Tseytlin for helpful discussions.
We also thank D.~O'Connell for important comments
on an earlier version of this paper.  This research was supported by the
US Department of Energy under contracts DE--FG03--91ER40662 and
DE-SC0007859.  The work of S.D. was supported by a US Department of
Energy Graduate Student Fellowship under contract DE-SC0008279.

\appendix

\section{Two-Loop Dimensionally Regularized Integrals}
\label{sec:DimRegAppendix}

In this appendix, we explicitly compute the divergent parts of dimensionally
regularized two-loop
integrals in $D=4-2\epsilon$, appearing in Section~\ref{sec:twoLoopUV}.
In general, both ultraviolet and
infrared divergences appear as poles in $\epsilon$ so we must
subtract the infrared ones in order to obtain the ultraviolet ones.

We start with the planar double-box integral, displayed in
\fig{Fig:PandNP}(a), following the discussion in Section 4 of
  Ref.~\cite{BernDeFreitasDixon},
\begin{eqnarray}
&& \hskip -.3 cm 
\mathcal{I}_4^{\mathrm{P}}[\mathcal{P}(\lambda_i,p,q,k_i)](s,t) \nonumber \\
&&  \hskip -.3 cm 
\hphantom{\mathcal{I}_4^{\mathrm{P}}[\mathcal{P}}\equiv\int\frac{d^Dp}{(2\pi)^D}\frac{d^Dq}{(2\pi)^D}\frac{\mathcal{P}(\lambda_i,p,q,k_i)}{p^2q^2(p+q)^2(p-k_1)^2(p-k_1-k_2)^2(q-k_4)^2(q-k_3-k_4)^2}\,. \hskip 1 cm 
\end{eqnarray}
 Using Schwinger parameters, we rewrite the planar double-box
integral with constant numerator as
\begin{equation}
 \I_4^{\mathrm{P}} [1] (s,t)=
\frac{1}{(4\pi)^D}\prod^7_{i=1}\int_0^\infty dt_{i} \left[\Delta_\P(T)\right]^{-\frac{D}{2}}\mathrm{exp}\left[ -\frac{ Q_\P(s,t,t_i)}{\Delta_\P(T)}\right],
\label{eq:schwinger}
\end{equation}
where
\begin{equation}
 \Delta_\P(T) = (T_pT_q+T_pT_{pq}+T_qT_{pq})\,,
\label{eq:planarDelta}
\end{equation}
and
\begin{equation}
T_p = t_3+t_4+t_5\,,\;\;\;T_q=t_1+t_2+t_7\,,\;\;\;T_{pq}=t_6 \,. 
\end{equation}
$T_p$, $T_q$, and $T_{pq}$ are sums of Schwinger parameters corresponding to propagators with loop momenta $p$, $q$, and $p+q$, respectively.  We also have
\begin{equation}
  Q_\P(s,t,t_i) = -s \, \bigg( t_1 t_2 T_p + t_3 t_4 T_q 
                       + t_6 (t_1+t_3)(t_2+t_4) \bigg) 
              -t \, t_5 t_6 t_7 \,. 
\label{Qnumer}
\end{equation}
To account for factors of $\lambda_p^2$, $\lambda_q^2$, and $\lambda_{p+q}^2$ in the numerator, we take derivatives on the $(-2\epsilon)$-dimensional part of the (Wick-rotated) integral:
\begin{equation}
\int d\lambda_p^{-2\epsilon}d\lambda_q^{-2\epsilon}\mathrm{exp}\left[ -T_p\lambda^2_p-T_q\lambda^2_q-T_{pq}\lambda^2_{p+q}\right]\propto \left[\Delta_\P(T)\right]^{\epsilon}\,,
\end{equation}
with respect to $T_p$, $T_q$, and $T_{pq}$.  This introduces additional factors to be inserted in the integrand in Eq.~\eqref{eq:schwinger}.  For example,
\begin{eqnarray}
(\lambda_p^2)^4 & \rightarrow& -\epsilon(1-\epsilon)(2-\epsilon)(3-\epsilon)\left(\frac{T_q+T_{pq}}{\Delta_\P(T)}\right)^4, \nonumber \\
(\lambda_p^2)^3\lambda_q^2 & \rightarrow & \epsilon^2(1-\epsilon)(2-\epsilon)\frac{(T_q+T_{pq})^2}{\Delta_\P(T)^3}-\epsilon(1-\epsilon)(2-\epsilon)(3-\epsilon)\frac{(T_q+T_{pq})^2T_{pq}^2}{\Delta_\P(T)^4}\,,\nonumber\\
(\lambda_p^2)^2\lambda_q^2\lambda_{p+q}^2 \hspace{-.1cm} & \rightarrow &\epsilon^2(1-\epsilon)^2\frac{1}{\Delta_\P(T)^2}+\epsilon(1-\epsilon)(2-\epsilon)\frac{\epsilon(T_q^2+T_{pq}^2)+2T_qT_{pq}}{\Delta_\P(T)^3} \label{eq:extra} \\
&& \hspace{6cm} \null 
-\epsilon(1-\epsilon)(2-\epsilon)(3-\epsilon)\frac{T_q^2T_{pq}^2}{\Delta_\P(T)^4}\,. \nonumber \hskip .5 cm 
\end{eqnarray}
We account for extra factors of $\Delta_\P^a(T)$ by shifting the
dimension $D\rightarrow D-2a$.  Following Smirnov~\cite{Smirnov}, we
change six of the seven Schwinger parameters to Feynman parameters
with the delta-function constraint $\sum_{i\neq 6}\alpha_i=1$:
\begin{equation}
\I_4^\P [{\cal P}  (\mud_p, \mud_q)] (s,t)=\frac{\Gamma[7-D+\gamma]}{(4\pi)^D}
\int_0^{\infty}\hspace{-.1cm}d\alpha_6\prod_{i\neq6}\int_0^1 \hspace{-.1cm}d\alpha_{i} \delta\biggl(\hspace{-.05cm}1-\sum_{i\neq6}\alpha_i\biggr)
\frac{\left[\Delta_\P(T)\right]^{7-\frac{3D}{2}+\gamma}}{ \left[ Q_\P(s,t,\alpha_i)\right]^{7-D+\gamma}}D(\alpha_i) \,,
\label{FeynPa}
\end{equation}
where $D(\alpha_i)$ represents the extra factors in one term of 
Eq.~\eqref{eq:extra}, with $t_i\rightarrow\alpha_i$.  The parameter
$\gamma$ counts the factors of $\alpha_i$ in $D(\alpha_i)$ and can
take on values 0, 2, and 4 for the integrals under consideration here.
Next we perform a change of variables
that imposes the delta-function constraint~\cite{Smirnov}:
\begin{eqnarray}
&&\alpha_1=\beta_1\xi_3\,, \hskip .6 cm 
\alpha_2=(1-\xi_5)(1-\xi_4)\,,     \hskip .6 cm 
\alpha_3 =\beta_2\xi_1\,, \hskip .6 cm
\alpha_4=\xi_5(1-\xi_2)\,, \nonumber \\
&&\alpha_5=\beta_2(1-\xi_1)\,, \hskip .6cm
\alpha_7=\beta_1(1-\xi_3)\,, \hskip .6cm
\beta_1=(1-\xi_5)\xi_4\,, \hskip .6cm
\beta_2=\xi_5\xi_2\,.
\end{eqnarray}
We then integrate these parameters to obtain a Mellin-Barnes representation, which we again integrate.  Finally we arrive at the dimensionally regularized results of our required planar double-box integrals:
\begin{align}
\mathcal{I}_4^{\mathrm{P}}[(\lambda_p^2)^4](s,t)&=\mathcal{I'}^{\mathrm{P}}-\frac{1}{(4\pi)^4}\frac{s+2t}{360\epsilon}+\mathcal{O}(\epsilon^0)\,,  \nonumber \\
\mathcal{I}_4^{\mathrm{P}}[(\lambda_{p+q}^2)^4](s,t)&=2\mathcal{I'}^{\mathrm{P}}-\frac{1}{(4\pi)^4}\frac{29s+4t}{180\epsilon}+\mathcal{O}(\epsilon^0)\,, \nonumber \\
\mathcal{I}_4^{\mathrm{P}}[(\lambda_p^2)^3\lambda_q^2](s,t)&=-\frac{1}{(4\pi)^4}\frac{s}{480\epsilon}+\mathcal{O}(\epsilon^0)\,, \nonumber \\
\mathcal{I}_4^{\mathrm{P}}[(\lambda_p^2)^3\lambda_{p+q}^2](s,t)&=\mathcal{I'}^{\mathrm{P}}+\frac{1}{(4\pi)^4}\frac{s-t}{360\epsilon}+\mathcal{O}(\epsilon^0)\,, \nonumber \\
\mathcal{I}_4^{\mathrm{P}}[(\lambda_p^2)^2(\lambda_q^2)^2](s,t)&=\mathcal{O}(\epsilon^0)\,, \nonumber \\
\mathcal{I}_4^{\mathrm{P}}[(\lambda_p^2)^2(\lambda_{p+q}^2)^2](s,t)&=\mathcal{I'}^{\mathrm{P}}-\frac{1}{(4\pi)^4}\frac{s+2t}{720\epsilon}+\mathcal{O}(\epsilon^0)\,, \nonumber \\
\mathcal{I}_4^{\mathrm{P}}[(\lambda_p^2)^2\lambda_q^2\lambda_{p+q}^2](s,t)&=\frac{1}{(4\pi)^4}\frac{s}{720\epsilon}+\mathcal{O}(\epsilon^0)\,, \nonumber \\
\mathcal{I}_4^{\mathrm{P}}[\lambda_p^2\lambda_q^2(\lambda_{p+q}^2)^2](s,t)&=-\frac{1}{(4\pi)^4}\frac{s}{240\epsilon}+\mathcal{O}(\epsilon^0)\,, \nonumber \\
\mathcal{I}_4^{\mathrm{P}}[\lambda_p^2(\lambda_{p+q}^2)^3](s,t)&=\mathcal{I'}^{\mathrm{P}}-\frac{1}{(4\pi)^4}\frac{5s+t}{180\epsilon}+\mathcal{O}(\epsilon^0)\,,
\label{eq:dimRegPlanar}
\end{align}
where
\begin{eqnarray}
\mathcal{I'}^\mathrm{P}&\equiv&\frac{1}{(4\pi)^4}
\left[\frac{1}{840s\epsilon^2}\left(2s^2+s t+2t^2\right)
 (-s)^{-2\epsilon}e^{-2\epsilon\gamma_{\mathrm{E}}}\right. \nonumber \\
&&\hspace{.8cm}\null
+\frac{1}{88200s u^4\epsilon}(4s^6+753s^5 t+4306s^4 t^2+9144s^3 t^3 \nonumber \\
&&\hspace{5cm}\null
-315\pi^2 s^3 t^3+9381 s^2 t^4+4813 s t^5+1019 t^6) \nonumber \\
&&\hspace{.8cm}\left.\null
+\frac{t^3(11s^2+7s t+2t^2)}{840s u^3\epsilon}
\log\Bigl(\frac{t}{s}\Bigr)
-\frac{s^2t^3}{280u^4\epsilon} \log^2\Bigl(\frac{t}{s}\Bigr)\right] 
+\mathcal{O}(\epsilon^0)\,.
\end{eqnarray}
All integrals above are symmetric under $\lambda_p\leftrightarrow\lambda_q$.

Next we look at the nonplanar double-box integrals:
\begin{align}
&\mathcal{I}_4^{\mathrm{NP}}[\mathcal{P}(\lambda_i,p,q,k_i)](s,t) \nonumber \\
&\hphantom{\mathcal{I}_4^{\mathrm{P}}[\mathcal{P}}\equiv\int\frac{d^Dp}{(2\pi)^D}\frac{d^Dq}{(2\pi)^D}\frac{\mathcal{P}(\lambda_i,p,q,k_i)}{p^2q^2(p+q)^2(p-k_1)^2(q-k_2)^2(p+q+k_3)^2(p+q+k_3+k_4)^2}\,,
\end{align}
whose evaluation follows that of the planar double-box integrals quite closely.  $\Delta_\NP(T)$ takes the same form as $\Delta_\P(T)$ in Eq.~\eqref{eq:planarDelta}, except that
\begin{equation}
T_p = t_1+t_2\,,\;\;\;T_q=t_3+t_4\,,\;\;\;T_{pq}=t_5+t_6+t_7 \,.
\end{equation}
We then also have
\begin{equation}
  Q_\NP(s,t,u,t_i)  = -s \, \bigl( t_1 t_3 t_5 + t_2 t_4 t_7 
              + t_5 t_7 (T_p+T_q) \bigr)
              - t \, t_2 t_3 t_6 -u \, t_1 t_4 t_6 \,. 
\end{equation}
In this case, we find it advantageous to only change the four Schwinger parameters associated with $T_p$ and $T_q$ to Feynman parameters, resulting in
\begin{eqnarray}
 \I_4^\NP [{\cal P}  (\mud_p, \mud_q)] &=&\frac{\Gamma[7-D+\gamma]}{(4\pi)^D} \\
&& \null \times 
\prod^7_{i=5}\int_0^\infty d\alpha_{i} \prod^4_{j=1}
\int_0^1 d\alpha_{j} \delta\left(1-\sum_{i=1}^4\alpha_i\right)
\frac{\left[\Delta_\NP(T)\right]^{7-\frac{3D}{2}+\gamma}}
     {\left[ Q_\NP(s,t,u,\alpha_i)\right]^{7-D+\gamma}}
 D(\alpha_i) \,. \nonumber \hskip .5 cm
\end{eqnarray}
We impose the delta-function constraint via further redefinition:
\begin{equation}
\alpha_1=\xi_3(1-\xi_1)\,, \hskip .8 cm 
\alpha_2=\xi_3\xi_1\,,     \hskip .8 cm 
\alpha_3 =(1-\xi_3)(1-\xi_2)\,, \hskip .8 cm
\alpha_4=(1-\xi_3)\xi_2\,.
\end{equation}
Once again we can straightforwardly integrate the parameters and use the Mellin-Barnes representation to evaluate our required nonplanar double-box integrals:
\begin{align}
\mathcal{I}_4^{\mathrm{NP}}[(\lambda_p^2)^4](s,t)&=\mathcal{I'}^{\mathrm{NP}}-\frac{1}{(4\pi)^4}\frac{215s^2+342st+342t^2}{50400s\epsilon}+\mathcal{O}(\epsilon^0)\,, \nonumber \\
\mathcal{I}_4^{\mathrm{NP}}[(\lambda_{p+q}^2)^4](s,t)&=\frac{1}{(4\pi)^4}\frac{s}{80\epsilon}+\mathcal{O}(\epsilon^0)\,, \nonumber \\
\mathcal{I}_4^{\mathrm{NP}}[(\lambda_p^2)^3\lambda_q^2](s,t)&=\mathcal{I'}^{\mathrm{NP}}-\frac{1}{(4\pi)^4}\frac{215s^2+342st+342t^2}{50400s\epsilon}+\mathcal{O}(\epsilon^0)\,, \nonumber \\
\mathcal{I}_4^{\mathrm{NP}}[(\lambda_p^2)^3\lambda_{p+q}^2](s,t)&=\mathcal{O}(\epsilon^0)\,, \nonumber \\
\mathcal{I}_4^{\mathrm{NP}}[(\lambda_p^2)^2(\lambda_q^2)^2](s,t)&=\mathcal{I'}^{\mathrm{NP}}-\frac{1}{(4\pi)^4}\frac{230s^2+171st+171t^2}{25200s\epsilon}+\mathcal{O}(\epsilon^0)\,, \nonumber \\
\mathcal{I}_4^{\mathrm{NP}}[(\lambda_p^2)^2(\lambda_{p+q}^2)^2](s,t)&=\frac{1}{(4\pi)^4}\frac{s}{160\epsilon}+\mathcal{O}(\epsilon^0)\,, \nonumber \\
\mathcal{I}_4^{\mathrm{NP}}[(\lambda_p^2)^2\lambda_q^2\lambda_{p+q}^2](s,t)&=\frac{1}{(4\pi)^4}\frac{s}{1440\epsilon}+\mathcal{O}(\epsilon^0)\,, \nonumber \\
\mathcal{I}_4^{\mathrm{NP}}[\lambda_p^2\lambda_q^2(\lambda_{p+q}^2)^2](s,t)&=\mathcal{O}(\epsilon^0)\,, \nonumber \\
\mathcal{I}_4^{\mathrm{NP}}[\lambda_p^2(\lambda_{p+q}^2)^3](s,t)&=\frac{1}{(4\pi)^4}\frac{s}{160\epsilon}+\mathcal{O}(\epsilon^0)\,,
\label{eq:dimRegNonplanar}
\end{align}
where
\begin{eqnarray}
\mathcal{I'}^\mathrm{NP}& \equiv&\frac{1}{(4\pi)^4}\left[\frac{1}{840s\epsilon^2}\left(2t^2+tu+2u^2\right)(-s)^{-\epsilon}(-t)^{-\epsilon}e^{-2\epsilon\gamma_{\mathrm{E}}}\right. \nonumber \\
&&\hspace{1.2cm} \null
+\frac{1}{352800s^5\epsilon}
     (5581u^6+25188u^5t+51783u^4t^2+64352u^3t^3 \nonumber \\
&&\hspace{5cm}\null
-1260\pi^2u^3t^3+51783u^2t^4+25188ut^5+5581t^6) \nonumber \\
&&\hspace{1.2cm} \left. \null
+\frac{u^3(11t^2+7tu+2u^2)}{840s^4\epsilon}\log\Bigl(\frac{u}{t}\Bigr)
-\frac{t^3u^3}{280s^5\epsilon} \log^2\Bigl(\frac{u}{t}\Bigr)\right]+\mathcal{O}(\epsilon^0) \,.
\end{eqnarray}
As with the planar results, the above are valid under the exchange
$\lambda_p\leftrightarrow\lambda_q$.

Finally we evaluate the bow-tie integrals:
\begin{align}
&\mathcal{I}_4^{\mathrm{bow\hbox{-}tie}}[\mathcal{P}(\lambda_i,p,q,k_i)](s) \nonumber \\
&\hphantom{\mathcal{I}_4^{\mathrm{P}}[\mathcal{P}]}\equiv\int\frac{d^Dp}{(2\pi)^D}\frac{d^Dq}{(2\pi)^D}\frac{\mathcal{P}(\lambda_i,p,q,k_i)}{p^2q^2(p-k_1)^2(p-k_1-k_2)^2(q-k_4)^2(q-k_3-k_4)^2}\,.
\end{align}
The bow-tie integrals are relatively simple because they are products of two one-loop integrals.  Similar techniques involving Schwinger parameters and Mellin-Barnes representations can be used on each one-loop integral.  Since bubbles with a massless leg vanish in dimensional regularization, the replacement $(p+q)^2\rightarrow 2p\cdot q$ is valid in the numerator.  We also use the tensor reduction $(\lambda_p\cdot\lambda_q)^2\rightarrow\lambda_p^2\lambda_q^2/(-2\epsilon)$.  For the bow-tie integrals appearing in Eq.~\eqref{eq:twoLoopDoubleCopy}, this tensor reduction is the only source of an ultraviolet divergence.  When evaluating the bow-tie contributions then, we expose $(\lambda_p\cdot\lambda_q)^2$ factors through the substitutions,
\begin{align}
&\lambda_{p+q}^2\rightarrow \lambda_p^2+\lambda_q^2+2(\lambda_p\cdot\lambda_q)\,,
&(p+q)^2\rightarrow (2p_{(4)}\cdot q_{(4)})-2(\lambda_p\cdot\lambda_q)\,.
\end{align}
Only terms containing a $(\lambda_p\cdot\lambda_q)^2$ are ultraviolet divergent; there are no terms with $(\lambda_p\cdot\lambda_q)^4$ or higher powers of $(\lambda_p\cdot\lambda_q)$.  The relevant bow-tie integrals are then given by
\begin{align}
\mathcal{I}_4^{\mathrm{bow\hbox{-}tie}}[(\lambda_p^2)^2(\lambda_p\cdot\lambda_q)^2](s)&=\frac{1}{(4\pi)^4}\frac{s^2}{720\epsilon} +\mathcal{O}(\epsilon^0)\,, \nonumber \\
\mathcal{I}_4^{\mathrm{bow\hbox{-}tie}}[\lambda_p^2\lambda_q^2(\lambda_p\cdot\lambda_q)^2](s)&=\frac{1}{(4\pi)^4}\frac{s^2}{1152\epsilon}+\mathcal{O}(\epsilon^0)\,, \nonumber \\
\mathcal{I}_4^{\mathrm{bow\hbox{-}tie}}[(\lambda_p^2)^2\lambda_q^2(\lambda_p\cdot\lambda_q)^2](s)&=\frac{1}{(4\pi)^4}\frac{s^3}{8640\epsilon}+\mathcal{O}(\epsilon^0)\,, \nonumber \\
\mathcal{I}_4^{\mathrm{bow\hbox{-}tie}}[(\lambda_p^2)^2(\lambda_q^2)^2(\lambda_p\cdot\lambda_q)^2](s)&=\frac{1}{(4\pi)^4}\frac{s^4}{64800\epsilon}+\mathcal{O}(\epsilon^0)\,, \nonumber \\
\mathcal{I}_4^{\mathrm{bow\hbox{-}tie}}[(\lambda_p^2)^2(\lambda_p\cdot\lambda_q)^2(2p_{(4)}\cdot q_{(4)})](s,t)&=-\frac{1}{(4\pi)^4}\frac{s^2(10s-t)}{15120\epsilon}+\mathcal{O}(\epsilon^0)\,, \nonumber \\
\mathcal{I}_4^{\mathrm{bow\hbox{-}tie}}[\lambda_p^2\lambda_q^2(\lambda_p\cdot\lambda_q)^2(2p_{(4)}\cdot q_{(4)})](s,t)&=-\frac{1}{(4\pi)^4}\frac{s^2(12s-t)}{28800\epsilon}+\mathcal{O}(\epsilon^0)\,.
\label{eq:bowTieInts}
\end{align}
These are also symmetric under the exchange $\lambda_p\leftrightarrow\lambda_q$.

\section{Two-Loop Infrared Divergence}
\label{sec:IRAppendix}

In this appendix we obtain the two-loop infrared divergence for the
four-point all-plus-helicity graviton amplitude in the theory of
gravity coupled to a dilaton and an antisymmetric tensor using
dimensional regularization in $D=4-2\epsilon$.  We subtract the
infrared divergence from the total divergence to obtain the
ultraviolet divergence.  Infrared divergences in gravity can be
obtained by exponentiating the divergence found at the one-loop
order~\cite{WeinbergIR, SchnitzerIR,White}.  In the cases where there
is a divergence at one loop, the infrared singularities are `one-loop
exact'; however, in the all-plus-helicity gravitons case, the first
divergence occurs at two loops.  Nevertheless, the same principles
apply. More specifically we are concerned with the exponentiation of
the gravitational soft function, which describes the effects of soft
graviton exchange between external particles.

Following the discussion of Ref.~\cite{SchnitzerIR}, a gravity
scattering amplitude can be written as
\begin{align}
\mathcal{M}_n=S_n\cdot H_n \,,
\label{eq:ampParts}
\end{align}
where $S_n$ is the infrared-divergent soft function and $H_n$ is the infrared-finite hard function.  Each quantity in Eq.~\eqref{eq:ampParts} can be written as a loop expansion in powers of $(\kappa/2)^2(4\pi e^{-\gamma_E})^{\epsilon}$:
\begin{align}
\mathcal{M}_n=\sum_{L=0}^{\infty}\mathcal{M}_n^{(L)}, \hspace{1cm}S_n=1+\sum_{L=1}^{\infty}S_n^{(L)},\hspace{1cm}H_n=\sum_{L=0}^{\infty}H_n^{(L)}.
\end{align}
The soft function is given by the exponential of the lowest-order infrared divergence:
\begin{align}
S_n=\mathrm{exp}\left[\frac{\sigma_n}{\epsilon}\right],\hspace{.8cm}\sigma_n=\left(\frac{\kappa}{2}\right)^2\frac{1}{(4\pi)^{2-\epsilon}}e^{-\gamma_E\epsilon}\sum_{j=1}^n\sum_{i<j}s_{ij}\mathrm{log}\left(\frac{-s_{ij}}{\mu^2}\right), \hspace{.8cm}s_{ij}=(k_i+k_j)^2.
\end{align}
An $L$-loop amplitude can then be written as
\begin{align}
\mathcal{M}_n^{(L)}=\sum_{l=0}^L\frac{1}{(L-l)!}\left[\frac{\sigma_n}{\epsilon}\right]^{L-l}H_n^{(l)}(\epsilon)\,.
\end{align}
For four-point amplitudes, we have
\begin{align}
\sigma_4=\left(\frac{\kappa}{2}\right)^2\frac{2}{(4\pi)^{2-\epsilon}}e^{-\gamma_E\epsilon}\left[s\,\mathrm{log}\left(\frac{-s}{\mu^2}\right)+t\,\mathrm{log}\left(\frac{-t}{\mu^2}\right)+u\,\mathrm{log}\left(\frac{-u}{\mu^2}\right)\right],
\end{align}
and the one-loop infrared divergence is given by
\begin{align}
\left. \mathcal{M}_4^{(1)}\right|_{\mathrm{IR~div.}}
=\frac{\sigma_4}{\epsilon} \mathcal{M}_4^{(0)}.
\end{align}
We used this to subtract the infrared divergence from our
dimensionally regularized one-loop result in
Section~\ref{sec:OneLoop4D} to isolate the ultraviolet divergence.
The four-point two-loop infrared divergence is given by
\begin{align}
\left. \mathcal{M}_4^{(2)}\right|_{\mathrm{IR~div.}}= \frac{1}{2}\left[\frac{\sigma_4}{\epsilon}\right]^2\mathcal{M}_4^{(0)}+\left.\frac{\sigma_4}{\epsilon}H_4^{(1)}(\epsilon)\right|_{\mathrm{IR~div.}}.
\end{align}
For the all-plus-helicity gravitons case, the tree amplitude $\mathcal{M}_4^{(0)}$ vanishes.  The one-loop amplitude is therefore infrared finite and equal to the one-loop infrared-finite hard function.  The one-loop amplitude can be computed using the double-copy procedure in Section~\ref{sec:OneLoop4D} and is given by~\cite{FourPointAllplusGrav}
\begin{align}
\mathcal{M}^{(1)}(1^+,2^+,3^+,4^+)=-\left(\frac{\kappa}{2}\right)^4\frac{i}{(4\pi)^2}\left(\frac{[1\,2][3\,4]}{\langle 1\,2\rangle\langle 3\,4\rangle}\right)^2\frac{(D_s-2)^2}{240}\left(s^2+t^2+u^2\right).
\end{align}
The two-loop infrared divergence is then 
\begin{align}
\left.\mathcal{M}^{(2)}(1^+,2^+,3^+,4^+)\right|_{\mathrm{IR~div.}}=&-\frac{1}{\epsilon}\left(\frac{\kappa}{2}\right)^6\frac{i}{(4\pi)^4}\left(\frac{[1\,2][3\,4]}{\langle 1\,2\rangle\langle 3\,4\rangle}\right)^2\frac{(D_s-2)^2}{120}\left(s^2+t^2+u^2\right) \nonumber \\
&\hspace{.5cm}\times\left[s\,\mathrm{log}\left(\frac{-s}{\mu^2}\right)+t\,\mathrm{log}\left(\frac{-t}{\mu^2}\right)+u\,\mathrm{log}\left(\frac{-u}{\mu^2}\right)\right].
\end{align}

\section{Two-Loop Ultraviolet Divergences from Vacuum Integrals}
\label{sec:UVIntegralsAppendix}

In this appendix we compute the ultraviolet divergences of the integrals in Section~\ref{sec:twoLoopUV}.  The techniques are very similar to those used to study the one-loop ultraviolet properties of gravity in Section~\ref{sec:LoopUV}.  However, before we can use them, we must deal with the $(-2\epsilon)$-dimensional components $\lambda_p$, $\lambda_q$, and $\lambda_{p+q}$ in the numerators of the integrals using the techniques in Section 4.1 of Ref.~\cite{BernDeFreitasDixon}.

The effect of inserting factors of $\lambda_p$, $\lambda_q$, and $\lambda_{p+q}$ into the planar and nonplanar double-box integrals is very similar to inserting factors of $v\cdot p$, $v\cdot q$, and $v\cdot(p+q)$, where
\begin{align}
v^{\mu}\equiv\epsilon^{\mu}_{\hphantom{\mu}\nu_1\nu_2\nu_3}k_1^{\nu_1}k_2^{\nu_2}k_3^{\nu_3}\,.
\end{align}
Example parameter insertions for factors of $\lambda_i$ are given in Eq.~\eqref{eq:extra}.  For polynomials in $v\cdot p$ and $v\cdot q$, we have
\begin{align}
(v\cdot p)^8\rightarrow&105\left(\frac{stu}{8}\right)^4\frac{(T_q+T_{pq})^4}{\Delta^4}\,, \nonumber \\
(v\cdot p)^6(v\cdot q)^2\rightarrow&\left(\frac{stu}{8}\right)^4\left[15\frac{(T_q+T_{pq})^2}{\Delta^3}+105\frac{(T_q+T_{pq})^2T_{pq}^2}{\Delta^4}\right], \nonumber \\
(v\cdot p)^4(v\cdot q)^4\rightarrow&\left(\frac{stu}{8}\right)^4\left[9\frac{1}{\Delta^2}+90\frac{T_{pq}^2}{\Delta^3}+105\frac{T_{pq}^4}{\Delta^4}\right], \nonumber \\
(v\cdot p)^4(v\cdot q)^2(v\cdot(p+q))^2\rightarrow&\left(\frac{stu}{8}\right)^4\left[9\frac{1}{\Delta^2}+15\frac{3T_q^2+3T_{pq}^2-2(T_q+T_{pq})^2}{\Delta^3}+105\frac{T_q^2T_{pq}^2}{\Delta^4}\right].
\label{eq:extraV}
\end{align}
These are valid for both the planar and nonplanar double boxes
provided the corresponding definitions for $\Delta$, $T_p$, $T_q$, and
$T_{pq}$ given in \app{sec:DimRegAppendix} are used.

We can also relate polynomials in $v\cdot p$ and $v\cdot q$ to the
$\lambda_i$.  The four-dimensional component of the loop momenta $p$
can be written as
\begin{align}
p^{\mu}_{[4]}\equiv c_1^p k_1^{\mu}+c_2^p k_2^{\mu}+c_3^p k_3^{\mu}+c_v^p v^{\mu},
\end{align}
where
\begin{align}
c_1^p=&\frac{1}{2su}\left[-t(2p\cdot k_1)+u(2p\cdot k_2)+s(2p\cdot k_3)\right], \nonumber \\
c_2^p=&\frac{1}{2st}\left[t(2p\cdot k_1)-u(2p\cdot k_2)+s(2p\cdot k_3)\right], \nonumber \\
c_3^p=&\frac{1}{2tu}\left[t(2p\cdot k_1)+u(2p\cdot k_2)-s(2p\cdot k_3)\right], \nonumber \\
c_v^p=&-\frac{4}{stu}\epsilon_{\mu\nu_1\nu_2\nu_3}p^{\mu}k_1^{\nu_1}k_2^{\nu_2}k_3^{\nu_3}=-\frac{4}{stu}v\cdot p\,.
\end{align}
We therefore have
\begin{align}
p^2+\lambda_p^2=p_{[4]}\cdot p_{[4]}=sc_1^pc_2^p+tc_2^pc_3^p+uc_1^pc_3^p-\frac{1}{4}stu(c_v^p)^2\,,
\end{align}
or
\begin{align}
\lambda_p^2=-\frac{4}{stu}(v\cdot p)^2+\hat{\mathcal{P}}_p\,,
\end{align}
where
\begin{align}
\hat{\mathcal{P}}_p\equiv&-p^2+sc_1^pc_2^p+tc_2^pc_3^p+uc_1^pc_3^p\,.
\end{align}
Similarly, we have
\begin{align}
\lambda_q^2&=-\frac{4}{stu}(v\cdot q)^2+\hat{\mathcal{P}}_q\,, \nonumber \\
\lambda_{p+q}^2&=-\frac{4}{stu}(v\cdot (p + q))^2+\hat{\mathcal{P}}_{pq}\,,
\end{align}
where
\begin{align}
\hat{\mathcal{P}}_q\equiv&-q^2+s\,c_1^qc_2^q+t\,c_2^qc_3^q+u\,c_1^qc_3^q\,, \nonumber \\
\hat{\mathcal{P}}_{pq}\equiv&-(p+q)^2+s(c_1^p+c_1^q)(c_2^p+c_2^q)+t(c_2^p+c_2^q)(c_3^p+c_3^q)+u(c_1^p+c_1^q)(c_3^p+c_3^q)\,.
\end{align}
These relations, along with the parameter replacements in Eqs.~\eqref{eq:extra}, \eqref{eq:extraV}, allow us to rewrite the integrals involving factors $\lambda_i$ in terms of integrals involving tensor products between the loop momenta and the external momenta.  For a general function $f(p\cdot k_i,q\cdot k_i)$, we have
\begin{align}
\int(\lambda_p^2)^4f=&-\frac{\epsilon(1-\epsilon)(2-\epsilon)(3-\epsilon)}{105}\left(\frac{8}{stu}\right)^4\int(v\cdot p)^8f \nonumber \\
=&-\frac{16\epsilon(1-\epsilon)(2-\epsilon)(3-\epsilon)}{(1-2\epsilon)(3-2\epsilon)(5-2\epsilon)(7-2\epsilon)}\int\hat{\mathcal{P}}_p^4f\,, \nonumber \\
\int(\lambda_p^2)^3\lambda_q^2f=&-\frac{16\epsilon(1-\epsilon)(2-\epsilon)(3-\epsilon)}{(1-2\epsilon)(3-2\epsilon)(5-2\epsilon)(7-2\epsilon)}\int\hat{\mathcal{P}}_p^3\hat{\mathcal{P}}_qf \nonumber \\
&+\frac{12\epsilon(1-\epsilon)(2-\epsilon)}{(3-2\epsilon)(5-2\epsilon)(7-2\epsilon)}\int\frac{\hat{\mathcal{P}}_p^2f}{\Delta}\,, \nonumber \\
\int(\lambda_p^2)^2(\lambda_q^2)^2f=&-\frac{16\epsilon(1-\epsilon)(2-\epsilon)(3-\epsilon)}{(1-2\epsilon)(3-2\epsilon)(5-2\epsilon)(7-2\epsilon)}\int\hat{\mathcal{P}}_p^2\hat{\mathcal{P}}_q^2f \nonumber \\
&+\frac{16\epsilon(1-\epsilon)(2-\epsilon)}{(3-2\epsilon)(5-2\epsilon)(7-2\epsilon)}\int\frac{\hat{\mathcal{P}}_p\hat{\mathcal{P}}_qf}{\Delta} \nonumber \\
&-\frac{6\epsilon(1-\epsilon)}{(5-2\epsilon)(7-2\epsilon)}\int\frac{f}{\Delta^2}\,, \nonumber \\
\int(\lambda_p^2)^2\lambda_q^2\lambda_{p+q}^2f=&-\frac{16\epsilon(1-\epsilon)(2-\epsilon)(3-\epsilon)}{(1-2\epsilon)(3-2\epsilon)(5-2\epsilon)(7-2\epsilon)}\int\hat{\mathcal{P}}_p^2\hat{\mathcal{P}}_q\hat{\mathcal{P}}_{pq}f \nonumber \\
&+\frac{4\epsilon(1-\epsilon)(2-\epsilon)}{(3-2\epsilon)(5-2\epsilon)(7-2\epsilon)}\int\frac{\hat{\mathcal{P}}_p(\hat{\mathcal{P}}_p+2\hat{\mathcal{P}}_q+2\hat{\mathcal{P}}_{pq})f}{\Delta} \nonumber \\
&-\frac{6\epsilon(1-\epsilon)}{(5-2\epsilon)(7-2\epsilon)}\int\frac{f}{\Delta^2}\,,
\label{eq:lambdaSub}
\end{align}
where a factor $1/\Delta$ indicates that a shift in dimension of the
integral should be made: $D\rightarrow D+2$,
$\epsilon\rightarrow\epsilon-1$ ($\epsilon$'s in prefactors in
Eq.~\eqref{eq:lambdaSub} should \emph{not} be shifted, however).

Once we have integrals in a form involving tensor products between the
loop momenta and external momenta, we expand in small external momenta
to reduce to logarithmically divergent integrals, just as we did in
the one-loop case.  This gives us vacuum integrals.  We then reduce
the tensors involving loop momenta using Lorentz covariance and insert
an infrared mass regulator. By integrating we obtain the ultraviolet
divergences. Since every prefactor in Eq.~\eqref{eq:lambdaSub}
contains a factor of $\epsilon$, to get the ultraviolet divergence, we
only need the $1/\epsilon^2$ pole of the integrals on the right-hand
side.  These leading contributions have no dependence on the mass
regulator, so we are unaffected by subdivergence issues due to the
mass regulator.  The ultraviolet divergences of the planar and
nonplanar double-box integrals are then
\begin{align}
\mathcal{I}_4^{\mathrm{P}}[(\lambda_p^2)^4](s,t)&=\mathcal{O}(\epsilon^0)\,,  \nonumber \\
\mathcal{I}_4^{\mathrm{P}}[(\lambda_{p+q}^2)^4](s,t)&=-\frac{1}{(4\pi)^4}\frac{14s+t}{90\epsilon}+\mathcal{O}(\epsilon^0)\,, \nonumber \\
\mathcal{I}_4^{\mathrm{P}}[(\lambda_p^2)^3\lambda_q^2](s,t)&=-\frac{1}{(4\pi)^4}\frac{s}{480\epsilon}+\mathcal{O}(\epsilon^0)\,, \nonumber \\
\mathcal{I}_4^{\mathrm{P}}[(\lambda_p^2)^3\lambda_{p+q}^2](s,t)&=\frac{1}{(4\pi)^4}\frac{2s+t}{360\epsilon}+\mathcal{O}(\epsilon^0)\,, \nonumber \\
\mathcal{I}_4^{\mathrm{P}}[(\lambda_p^2)^2(\lambda_q^2)^2](s,t)&=\mathcal{O}(\epsilon^0)\,, \nonumber \\
\mathcal{I}_4^{\mathrm{P}}[(\lambda_p^2)^2(\lambda_{p+q}^2)^2](s,t)&=\frac{1}{(4\pi)^4}\frac{s+2t}{720\epsilon}+\mathcal{O}(\epsilon^0)\,, \nonumber \\
\mathcal{I}_4^{\mathrm{P}}[(\lambda_p^2)^2\lambda_q^2\lambda_{p+q}^2](s,t)&=\frac{1}{(4\pi)^4}\frac{s}{720\epsilon}+\mathcal{O}(\epsilon^0)\,, \nonumber \\
\mathcal{I}_4^{\mathrm{P}}[\lambda_p^2\lambda_q^2(\lambda_{p+q}^2)^2](s,t)&=-\frac{1}{(4\pi)^4}\frac{s}{240\epsilon}+\mathcal{O}(\epsilon^0)\,, \nonumber \\
\mathcal{I}_4^{\mathrm{P}}[\lambda_p^2(\lambda_{p+q}^2)^3](s,t)&=-\frac{1}{(4\pi)^4}\frac{s}{40\epsilon}+\mathcal{O}(\epsilon^0)\,, \nonumber \\
\mathcal{I}_4^{\mathrm{NP}}[(\lambda_p^2)^4](s,t)&=-\frac{1}{(4\pi)^4}\frac{s}{80\epsilon}+\mathcal{O}(\epsilon^0)\,, \nonumber \\
\mathcal{I}_4^{\mathrm{NP}}[(\lambda_{p+q}^2)^4](s,t)&=-\frac{1}{(4\pi)^4}\frac{s}{80\epsilon}+\mathcal{O}(\epsilon^0)\,, \nonumber \\
\mathcal{I}_4^{\mathrm{NP}}[(\lambda_p^2)^3\lambda_q^2](s,t)&=\mathcal{O}(\epsilon^0)\,, \nonumber \\
\mathcal{I}_4^{\mathrm{NP}}[(\lambda_p^2)^3\lambda_{p+q}^2](s,t)&=-\frac{1}{(4\pi)^4}\frac{s}{80\epsilon}+\mathcal{O}(\epsilon^0)\,, \nonumber \\
\mathcal{I}_4^{\mathrm{NP}}[(\lambda_p^2)^2(\lambda_q^2)^2](s,t)&=-\frac{1}{(4\pi)^4}\frac{7s}{1440\epsilon}+\mathcal{O}(\epsilon^0)\,, \nonumber \\
\mathcal{I}_4^{\mathrm{NP}}[(\lambda_p^2)^2(\lambda_{p+q}^2)^2](s,t)&=-\frac{1}{(4\pi)^4}\frac{s}{160\epsilon}+\mathcal{O}(\epsilon^0)\,, \nonumber \\
\mathcal{I}_4^{\mathrm{NP}}[(\lambda_p^2)^2\lambda_q^2\lambda_{p+q}^2](s,t)&=\frac{1}{(4\pi)^4}\frac{s}{1440\epsilon}+\mathcal{O}(\epsilon^0)\,, \nonumber \\
\mathcal{I}_4^{\mathrm{NP}}[\lambda_p^2\lambda_q^2(\lambda_{p+q}^2)^2](s,t)&=\mathcal{O}(\epsilon^0)\,, \nonumber \\
\mathcal{I}_4^{\mathrm{NP}}[\lambda_p^2(\lambda_{p+q}^2)^3](s,t)&=-\frac{1}{(4\pi)^4}\frac{s}{160\epsilon}+\mathcal{O}(\epsilon^0)\,.
\end{align}
The bow-tie integrals do not contain infrared divergences, and their
ultraviolet divergences were computed in
Appendix~\ref{sec:DimRegAppendix}. Combining all the pieces then gives
us the ultraviolet divergence in \eqn{eq:twoLoopUV}.


\end{document}